\pdfoutput=1

\documentclass[preprint,5p,numbers,sort&compress]{elsarticle}

\usepackage{graphicx}
\usepackage{amssymb,amsmath}
\usepackage{natbib}
\usepackage{hyperref}
\hypersetup{pdftitle = The title of my PDF, pdfauthor = My name, pdfsubject= The subject, pdfkeywords = keyword1 keyword2 keyword3}
\hypersetup{colorlinks = true, linkcolor = blue, anchorcolor = red, citecolor = blue, filecolor = red, pagecolor = red, urlcolor = blue}

\journal{International Journal of Non-Linear Mechanics}

\newcommand{\norm}[1]{\left\lVert#1\right\rVert}

\begin{document}

\begin{frontmatter}
\title{Correlating the escape dynamics and the role of the normally hyperbolic invariant manifolds in a binary system of dwarf spheroidal galaxies}

\author[eez]{Euaggelos E. Zotos\corref{cor1}}
\ead{evzotos@physics.auth.gr}

\author[cj]{Christof Jung}

\cortext[cor1]{Corresponding author}

\address[eez]{Department of Physics, School of Science, \\
Aristotle University of Thessaloniki, \\
GR-541 24, Thessaloniki, Greece \\}

\address[cj]{Instituto de Ciencias F\'{i}sicas, \\
Universidad Nacional Aut\'{o}noma de M\'{e}xico \\
Av. Universidad s/n, 62251 Cuernavaca, Mexico}

\begin{abstract}
We elucidate the escape properties of stars moving in the combined gravitational field of a binary system of two interacting dwarf spheroidal galaxies. A galaxy model of three degrees of freedom is adopted for describing the dynamical properties of the Hamiltonian system. All the numerical values of the involved parameters are chosen having in mind the real binary system of the dwarf spheroidal galaxies NGC 147 and NGC 185. We distinguish between bounded (regular, sticky or chaotic) and escaping motion by classifying initial conditions of orbits in several types of two dimensional planes, considering only unbounded motion for several energy levels. We analyze the orbital structure of all types of two dimensional planes of initial conditions by locating the basins of escape and also by measuring the corresponding escape time of the orbits. Furthermore, the properties of the normally hyperbolic invariant manifolds (NHIMs), located in the vicinity of the index-1 saddle points $L_1$, $L_2$, and $L_3$, are also investigated. These manifolds are of great importance, as they control the flow of stars (between the two galaxies and toward the exterior region) over the different saddle points. In addition, bifurcation diagrams of the Lyapunov periodic orbits as well as restrictions of the Poincar\'e map to the NHIMs are presented for revealing the dynamics in the neighbourhood of the saddle points. Comparison between the current outcomes and previous related results is also made.
\end{abstract}


\end{frontmatter}

\section{Introduction}
\label{intro}

When we study a system of classical mechanics, then the index-1 saddles of the effective potential play a prominent role for the whole dynamics. They are the extremal points where the potential goes down along one direction only and goes up along all the other directions. For energies which are not too high over the saddle energy the escape from the inner potential holes to the outside regions or the transitions between various inner potential holes runs over these saddles. Usually we find some invariant subsets of the dynamics over the saddles and these sets are dynamically unstable in transverse directions, they are known as normally hyperbolic invariant manifolds (usual abbreviation NHIMs). General information on the NHIMs can be found in \cite{W94}. The states sitting over the potential saddles are also known under the name transition states (for their properties and connections to the NHIMs see \cite{WBW04,WBW05,WSW08}). If the saddle has index-1 and correspondingly the invariant subset has codimension 2 then the stable and the unstable manifolds of these NHIMs are of codimension 1 and they form impenetrable boundaries in the phase space and they delimit and channel the flow over the saddle. Generally, trajectories approach the saddle region along the stable manifolds and leave it again along the unstable manifolds of the NHIM. The projection of these stable and unstable manifolds from the phase space into the position space confines the saddle flow in the position space. As a consequence a major part of the study of any escape processes consists in the detailed investigation of the dynamics over the index-1 saddles of the effective potential. The saddle points of the effective potential in the rotating frame are also known as Lagrange points.

The most important periodic orbits within the NHIMs are the Lyapunov orbits over the saddle \cite{L07}. Therefore an important part of the investigation of the saddle dynamics consists of the study of the Lyapunov orbits and in particular of their development and bifurcation scenario as a function of the total energy. Also important for the global scenario are the most important periodic orbits growing out of the potential wells. For increasing values of the energy the orbits coming out of the potential minima frequently interact with the orbits coming from the saddles and in such cases they form a common development and bifurcation scenario.

In the present paper we restrict our consideration to systems with 3 degrees of freedom (3-dof systems). Then the Poincar\'{e} map acts on a 4-dimensional domain and its graphical representation needs some considerations. One attempt to display the full 4-dimensional map is the colour and rotation method (e.g., \cite{PZ14,KP11}), while another one is the use of slices (e.g., \cite{LRO14,RLBK14,OLKB16}). In the present paper we follow an alternative method. More precisely, we represent the saddle dynamics of 3-dof systems by intrinsically 2-dimensional Poincar\'{e} maps. This is possible along the following ideas: The relevant NHIMs in the full dimensional Poincar\'{e} map are invariant subsets of dimension 2 and the restriction of the Poincar\'{e} map to a NHIM exists and has as domain the 2 dimensional NHIM surface itself. So we can study this restricted map and it is the ideal tool to present the saddle dynamics and its development scenario by 2-dimensional graphics (e.g., \cite{GDJ14,GJ15}). It includes a graphical representation of the development scenario of the Lyapunov orbits and of some further important periodic orbits, related to the Lyapunov orbits.

One of the main purposes of the Poincar\'{e} map is to display the distribution of various types of motion in the phase space or in the energy shell. An alternative method to obtain this information consists in placing a dense grid of initial conditions on a general 2-dimensional surface in the phase space and determine the type of motion performed by the trajectory through each initial condition. In contrast to the basic idea of the Poincar\'{e} map the grid method does not need any return of the trajectory to the surface of initial conditions. As a consequence this implies an additional advantage of the grid method over the usual Poincar\'{e} map, namely that the grid method can keep track of escaping trajectories and of the exit chosen by them whereas these trajectories disappear from the domain of the usual Poincar\'{e} maps. This makes the grid method especially appropriate for escape systems.

So far little is known about the possible development scenarios of the saddle dynamics and in particular of the NHIMs (for some information on bifurcations of NHIMs see \cite{AB12,MCE13,MS14,LST06,TTK11,TTK15,TTT15}). At the moment, a general and complete theory for the bifurcations of the NHIMs in analogy to the bifurcation theory of fixed points seems to be out of reach. It is not yet possible to place the examples observed and analysed so far into some general frame work. Therefore, a central aspect of our entire programme of research consists in the collection of possible NHIM development scenarios. This implies to look for repeating patterns in the examples and to try to initiate along this route some kind of heuristic classification of the cases observed and to relate the various appearing cases with particular properties of the corresponding potential saddles.

A class of systems providing many fascinating examples of escape scenarios over saddles of the effective potential are systems of celestial mechanics treated in the rotating frame of reference, where the study of the dynamics around the Lagrange points has a long tradition. To keep track of similarities and differences of the various development possibilities we have to treat the collection of example systems by similar methods and to recognise well the different scenarios presented by these examples. Interestingly, we have already seen essentially different scenarios even though the various examples belong to a rather restricted class of systems. In the present paper we use as a further example system a rather simple model for two spheroidal dwarf galaxies in a bound state and rotating around their common center of mass. We study the dynamics of a test particle in the effective potential provided by this pair of galaxies. This system contains various index-1 saddles and one of them leads to a new kind of scenario which we did not yet see in the other examples studied so far (e.g., a barred galaxy \cite{JZ16} and a tidally limited star cluster \cite{ZJ17}). An important step in the present paper is a preliminary attempt to classify the scenarios observed so far and to relate them with other properties of the dynamics around the respective saddle point. In our choice of the free parameters of the system we have the galaxy pair NGC 147 and NGC 185 in mind.

Besides its interest for the scenarios of the escape dynamics this system is also interesting from the astronomical point of view. Dwarf spheroidal galaxies are usually found around our Galaxy and they are very faint with extremely low luminosity (e.g., \cite{M98}). Furthermore, the origin of this type of galaxies, in the Local Group, is still a long-standing subject of debate. It is believed that in the known universe they belong to the stellar objects which are most heavily dominated by dark matter (e.g., \cite{L09,WMO09}). Another intriguing aspect is the fact that dwarf spheroidal galaxies are poor in gas or even completely devoid of gas (e.g., \cite{GP09}). Nevertheless, they are characterized by pressure-supported spheroidal stellar components (e.g., \cite{M98}) as well as by a wide diversity of star formation histories (e.g., \cite{G00}).

Today dwarf spheroidal galaxies provide, due to their large velocity dispersions, a unique and excellent tool for testing the presence as well as the distribution of dark matter (for more details see the review of \cite{M97}). Observational data, regarding the stellar kinematics, allow us to interpret the existence, amount and of course the distribution of dark matter in dwarf spheroidal galaxies. More precisely we observe that stars move a lot faster than it would be normally expected due to the presence of the visible luminous matter only. Therefore, we need an extra amount of matter (apparently dark matter) in order to explain and justify the extra gravitational attraction. The same phenomenon, of fast moving stars, is also observed in disk galaxies (e.g., \cite{KS17}).

The motion of stars in binary galaxy systems can actually contribute to the observed velocity dispersions and particularly in the case of galaxies with low masses. This effect, usually known as ``binary contamination", plays a key role on the inferred amount of dark matter (e.g., \cite{MMB10}). This is true because if we incorrectly account for the motion of stars in binary systems, one could reach entirely false conclusions regarding the distribution of dark matter in dwarf spheroidal galaxies. At this point, it should be noted that the distribution of the dark matter in dwarf spheroidal galaxies is of paramount importance in testing several theories about galaxy formation in the universe.

The present paper belongs to a series of papers with common scope (numerical investigation of the escape dynamics and the role of the NHIMs in Hamiltonian stellar systems) and therefore the structure as well as the numerical techniques are similar to those used in the previous papers (e.g., \cite{JZ16,ZJ17}). The article is organized as follows: In Section \ref{mod} we present the binary galaxy model. In the following Section we conduct a thorough numerical investigation of the escape dynamics of stars. Section \ref{nhims} contains a detailed exploration of the dynamics in the vicinity of the saddle points and in particular of the NHIMs. In Section \ref{spr} we link the invariant manifolds with the stellar structures formed by escaping stars, while in Section \ref{disc} the conclusions are given. The paper ends with an appendix explaining the computation of the initial value of the momenta $p_y$.

\section{Description and properties of the binary galaxy model}
\label{mod}

Our binary stellar system consists of two dwarf spheroidal galaxies, with masses $M_1$ and $M_2$, which interact with each other by their mutual gravitational attraction. The third body (a star which acts as a test particle) moves in the combined gravitational field of the two galaxies. Both dwarf spheroidal galaxies move in a circular orbit in an inertial frame of reference Oxyz, with the origin located at the center of the mass of the system O, with a constant angular velocity $\Omega > 0$, given by Kepler's third law
\begin{equation}
\Omega = \sqrt{\frac{G M_{\rm t}}{R^3}},
\label{om}
\end{equation}
where $M_{\rm t} = M_1 + M_2$ is the total mass of the system, while $R$ is the distance between the centres of the two galaxies. At this point, we would like to emphasize that we assume that the two mass distributions of the galaxies are well
separated (the distance $R$ is sufficient), and therefore they behave as if they would be point masses. On this assumption Kepler's third law holds and the common angular velocity of the galaxies can be derived from Eq. (\ref{om}).

\begin{figure}[!t]
\begin{center}
\includegraphics[width=\hsize]{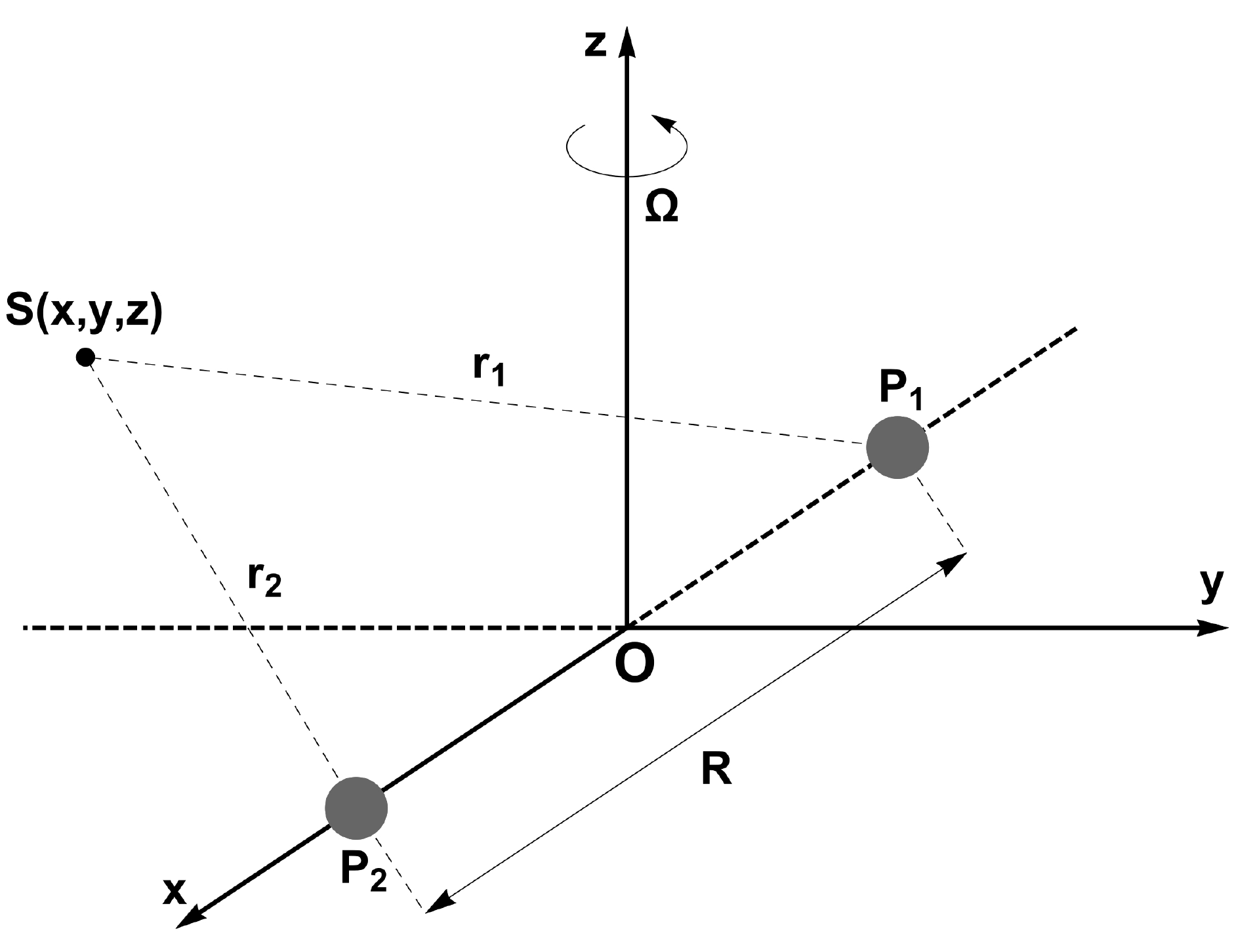}
\end{center}
\caption{A schematic plot depicting the configuration of the binary system of the two dwarf spheroidal galaxies.}
\label{sch}
\end{figure}

In the corotating coordinate system the centres of the two galaxies have fixed positions $P_1(x,y,z) = (x_1,0,0)$ and $P_2(x,y,z) = (x_2,0,0)$, where
\begin{equation}
x_1 = - \frac{M_2}{M_{\rm t}} R, \ \
x_2 = R - \frac{M_2}{M_{\rm t}} R = R + x_1.
\label{cents}
\end{equation}
In Fig. \ref{sch} we provide an illuminating schematic plot depicting the configuration of the binary system.

For modeling the dynamical properties of the dwarf spheroidal galaxies we shall use the spherically symmetric Plummer potential \cite{P11}. Therefore, the total gravitational potential, which is responsible for the motion of the test particle, is given by
\begin{equation}
\Phi_{\rm t}(x,y,z) = - \frac{G M_1}{r_1} - \frac{G M_2}{r_2},
\label{pot}
\end{equation}
where
\begin{align}
r_1 &= \sqrt{\left(x - x_1\right)^2 + y^2 + z^2 + c_1^2}, \nonumber\\
r_2 &= \sqrt{\left(x - x_2\right)^2 + y^2 + z^2 + c_2^2},
\label{dist}
\end{align}
are the distances of the test particle to the respective galaxies, while $c_1$ and $c_2$ are the core radii of the galaxies, which act as softening parameters.

The corresponding effective potential in the rotating frame of reference is
\begin{equation}
\Phi_{\rm eff}(x,y,z) = \Phi_{\rm t}(x,y,z) - \frac{\Omega^2}{2}\left(x^2 + y^2\right).
\label{eff}
\end{equation}
Looking carefully at Eq. (\ref{eff}) we see that it highly resembles the effective potential of the classical restricted three-body problem \cite{S67}. The only difference is the presence of the softening parameters $c_i$, $i = 1,2$. In fact, the presence of the softening parameters eliminates the problem of the critical collision orbits which is present in the classical restricted three-body problem. Furthermore, we may argue that the effective potential given in Eq. (\ref{eff}) corresponds to a generalized (softened) version of the classical restricted three-body problem. Here it should be noted that the effective potential had been used in \cite{Z16} for the investigation of the planar escape dynamics of a binary system of dwarf galaxies.

Then the equations of motion, according to the total gravitational potential $\Phi_{\rm t}$, read
\begin{align}
\dot{x} &= p_x + \Omega y, \nonumber \\
\dot{y} &= p_y - \Omega x, \nonumber \\
\dot{z} &= p_z, \nonumber \\
\dot{p_x} &= - \frac{\partial \Phi_{\rm t}}{\partial x} + \Omega p_y, \nonumber \\
\dot{p_y} &= - \frac{\partial \Phi_{\rm t}}{\partial y} - \Omega p_x, \nonumber \\
\dot{p_z} &= - \frac{\partial \Phi_{\rm t}}{\partial z},
\label{eqmot}
\end{align}
where, as usual, the dot indicates the derivative with respect to the time, while $p_x$, $p_y$ and $p_z$ are the canonical momenta per unit mass, conjugate to the coordinates $x$, $y$ and $z$, respectively.

For the calculation of standard chaos indicators we need to follow the time-evolution of deviation vectors ${\bf{w_i}}, i = 1,2$. Therefore the required set of variational equations is
\begin{align}
\dot{(\delta x)} &= \delta p_x + \Omega \delta y, \nonumber \\
\dot{(\delta y)} &= \delta p_y - \Omega \delta x, \nonumber \\
\dot{(\delta z)} &= \delta p_z, \nonumber \\
(\dot{\delta p_x}) &=
- \frac{\partial^2 \Phi_{\rm t}}{\partial x^2} \ \delta x
- \frac{\partial^2 \Phi_{\rm t}}{\partial x \partial y} \delta y
- \frac{\partial^2 \Phi_{\rm t}}{\partial x \partial z} \delta z + \Omega \delta p_y, \nonumber \\
(\dot{\delta p_y}) &=
- \frac{\partial^2 \Phi_{\rm t}}{\partial y \partial x} \delta x
- \frac{\partial^2 \Phi_{\rm t}}{\partial y^2} \delta y
- \frac{\partial^2 \Phi_{\rm t}}{\partial y \partial z} \delta z - \Omega \delta p_x, \nonumber \\
(\dot{\delta p_z}) &=
- \frac{\partial^2 \Phi_{\rm t}}{\partial z \partial x} \delta x
- \frac{\partial^2 \Phi_{\rm t}}{\partial z \partial y} \delta y
- \frac{\partial^2 \Phi_{\rm t}}{\partial z^2} \delta z.
\label{vareq}
\end{align}

The equations of motion (\ref{eqmot}) admit the following isolating integral, which governs the motion of a test particle (star) with a unit mass $(m = 1)$
\begin{equation}
H(x,y,z,p_x,p_y,p_z) = \frac{1}{2} \left(p_x^2 + p_y^2 + p_z^2 \right) + \Phi_{\rm t} - \Omega \ L_z = E,
\label{ham}
\end{equation}
where $E$ is the numerical value of the integral which is conserved, while $L_z = x \ p_y - y \ p_x$ is the angular momentum along the $z$ direction. Obviously, the Hamiltonian (\ref{ham}) is the well known energy integral.

The dynamical system and of course the equations of motion of Eq. (\ref{eqmot}) have two discrete symmetries. First, they are invariant under the transformation $z \to -z$, $p_z \to - p_z$ (i.e. under a reflection in the symmetry plane $z = 0$). In the special case of $M_1 = M_2$ and $c_1 = c_2$ we have the second and more interesting symmetry of a simultaneous inversion of the coordinates $(x,y,p_x,p_y)$ (i.e. the transformation $x \to - x$, $y \to - y$, $p_x \to - p_x$, and $p_y \to - p_y$). This transformation is equivalent to a rotation of the whole system around the $z$-axis by an angle $\pi$.

\begin{figure}[!t]
\begin{center}
\includegraphics[width=\hsize]{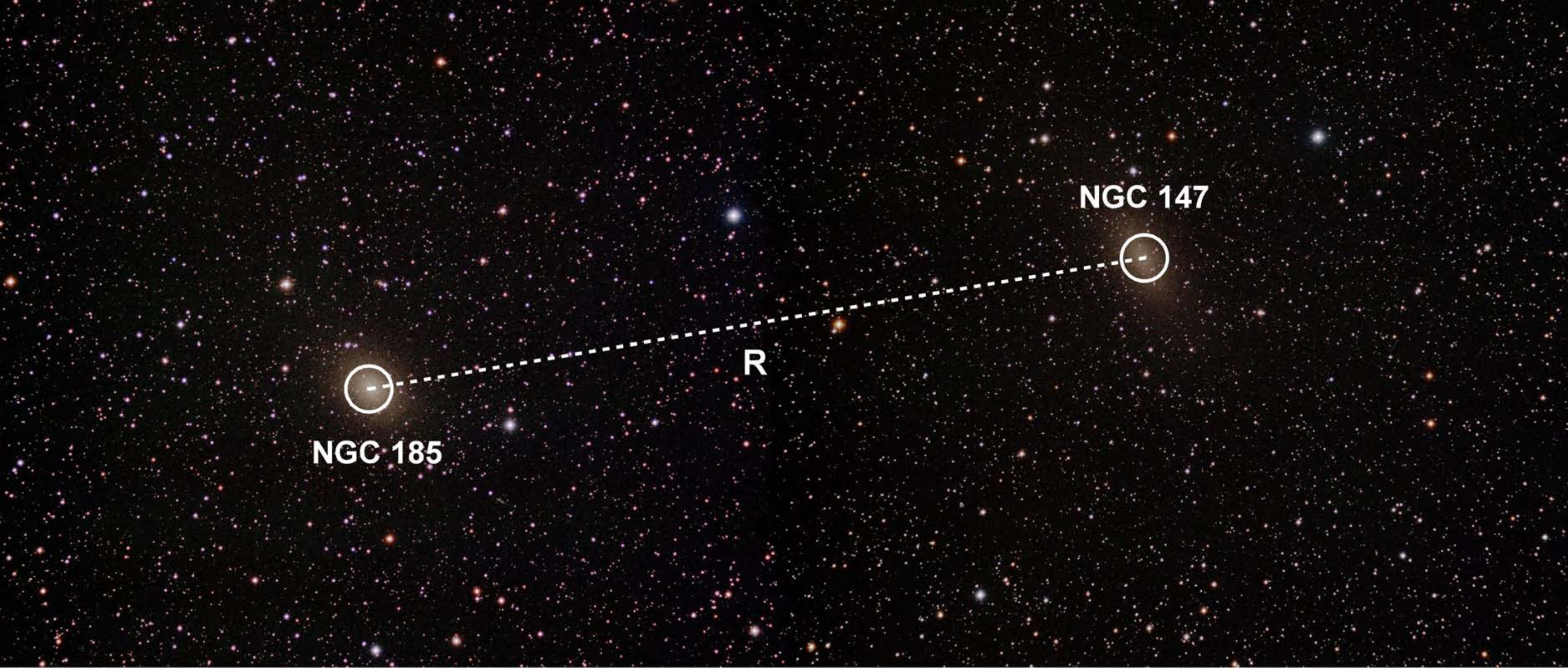}
\end{center}
\caption{A real image of the dwarf spheroidal galaxies NGC 147 and NGC 185 taken between 28 and 29 October of 2014, using an Orion 120mm EON Apochromatic Refractor telescope \citep{url}.}
\label{img}
\end{figure}

In our work we shall investigate the orbital dynamics of a binary galaxy system. All of the presently available observational data are consistent with the conclusion that NGC 147 and NGC 185 form a gravitationally bound system (see e.g., \cite{vdB98} and references therein). Both NGC 147 and NGC 185 are located in the constellation of Cassiopeia and they are members of the Local group of galaxies and they form a bound pair of satellite galaxies of the Andromeda Galaxy (M31). NGC 147 and NGC 185 are located in the sky about 6 degrees north of M31 and are separated by around 1 degree. NGC 185 is the brighter of the two galaxies (visual magnitude 10.1) and lies at a distance of around 2.05 $\times$ $10^6$ ly from us, while NGC 147 has a visual magnitude of 10.5 and lies at a distance of around 2.53 $\times$ $10^6$ ly. Fig. \ref{img} shows a real image of the two dwarf spheroidal galaxies.

We use a system of galactic units, where the unit of length is 10 kpc, the unit of mass is 2.22508 $\times 10^{10}$ M${_\odot}$ and the unit of time is 10$^{8}$ yr. The velocity unit is 97.86 km s$^{-1}$, the unit of angular momentum (per unit mass) is 978.6 km kpc s$^{-1}$, while $G$ is equal to unity. Finally, the energy unit (per unit mass) is 9.57 $\times$ $10^{3}$ km$^{2}$ s$^{-2}$. In these units the values of the involved parameters, for NGC 147 and NGC 185, are: $M_1 = M_2 = 0.1$, $c_1 = c_2 = 0.25$, while $R = 6$, according to \cite{AGF16} and \cite{GvdM10}. For simplicity in the numerical calculations we assume that both dwarf spheroidal galaxies NGC 147 and NGC 185 are identical, thus having the same mass as well as the same core radius.

\begin{figure}[!t]
\begin{center}
\includegraphics[width=\hsize]{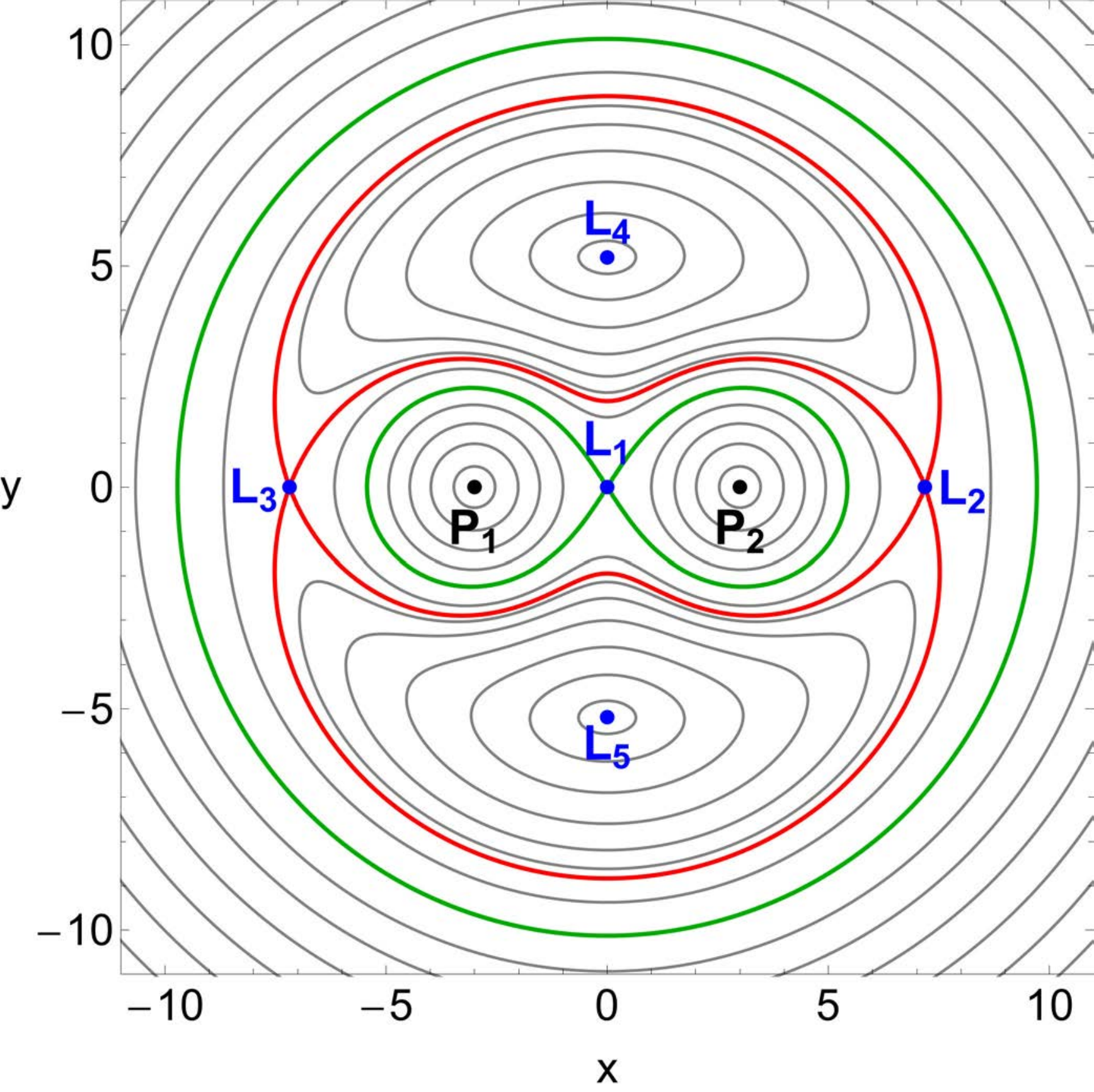}
\end{center}
\caption{The isoline contours of the effective potential $\Phi_{\rm eff}$ of the binary galaxy model on the configuration $(x,y)$ plane (when $z = 0$). The equipotential curves corresponding to the critical energy levels $E(L_1)$ and $E(L_2)$ are shown in green and red, respectively. The position of the five Lagrange points are indicated by blue dots, while black dots pinpoint the relative minima of the effective potential. (For the interpretation of references to colour in this figure caption and the corresponding text, the reader is referred to the electronic version of the article.)}
\label{conts}
\end{figure}

The effective potential of Eq. (\ref{eff}) has seven equilibrium points which fulfil the equation
\begin{equation}
\frac{\partial \Phi_{\rm eff}}{\partial x} = \frac{\partial \Phi_{\rm eff}}{\partial y} = \frac{\partial \Phi_{\rm eff}}{\partial z} = 0.
\label{lgs}
\end{equation}
There are two minima which we call $P_1$ and $P_2$ which almost coincide with the centres of the two galaxies. Next we have the three collinear points $L_1$, $L_2$ and $L_3$ located on the $x$-axis. Finally there are the two triangular points $L_4$ and $L_5$ with a nonzero $y$ coordinate (see Fig. \ref{conts}). The Lagrange points $L_j$ are very similar to the corresponding Lagrange points in the classical restricted three-body problem with pure point mass $1/r$ potentials. All collinear points are index-1 saddle points of the effective potential, while the stationary points $L_4$ and $L_5$ are index-2 saddle points of the effective potential. $L_1$ is located at the origin $(0,0)$, while the coordinates of $L_2$ and $L_3$ are $(x,y,z) = (\pm r_L,0,0)$, where $r_L = 7.18226801$ is the Lagrange radius (e.g., \cite{BT08}). On the other hand, $L_4$ and $L_5$ are located at $(x,y,z) = (0,\pm y^*,0)$, where $y^* = 5.19013487$.

The positions of the five Lagrange points are also indicated in the same figure. The numerical values of the effective potential at the saddle points are critical values of the Hamiltonian. In our case we have that $E(L_1) = -0.06643638$, $E(L_2) = E(L_3) = -0.05756783$, and $E(L_4) = E(L_5) = -0.04580439$. When $E > E(L_1)$, the channel near the origin opens thus allowing stars to circulate around both galaxies. Moreover when $E > E(L_2) = E(L_3)$ the zero velocity surfaces open and two symmetrical escape channels (exits) emerge in the vicinity of the Lagrange points $L_2$ and $L_3$. Through these channels, the stars are allowed to enter the exterior region (when $x < - r_L$ or when $x > + r_L$) and therefore are free to escape to infinity.

A double precision Bulirsch-Stoer \verb!FORTRAN 77! algorithm (e.g., \cite{PTVF92}) was used for integrating backwards and forwards the equations of motion (\ref{eqmot}) as well as the variational equations (\ref{vareq}). The time step of the numerical integration was of the order of $10^{-2}$ which is sufficient for the desired accuracy of our computations. Throughout our computations the numerical error regarding the conservation of the energy integral of motion given in Eq. (\ref{ham}) was smaller than $10^{-13}$, although there were cases that the corresponding error was smaller than $10^{-14}$. All graphical illustrations presented in this paper have been created using the latest version 11.2 of Mathematica$^{\circledR}$ (e.g., \cite{Wolf03}).

\section{Escape dynamics}
\label{esc}

In this Section we unveil the escape mechanism of stars moving in the gravitational field of the binary galaxy system. For this purpose we define, for several values $E$ of the energy integral, dense uniform grids of $1024 \times 1024$ initial conditions of orbits on the $(x,z)$ plane, always inside the corresponding energetically allowed area. Following a typical numerical approach, all orbits of the stars are launched with initial conditions inside the Lagrange radius $(x_0^2 + y_0^2 + z_0^2 \leq r_L^2)$. Furthermore, all orbits have pairs of initial conditions $(x_0,z_0)$, with $y_0 = p_{x0} = p_{z0} = 0$, while the initial value of $p_y$ is always obtained from the energy integral (\ref{ham}) as $p_{y_0} = p_y(x_0,y_0,z_0,p_{x_0},p_{z0},E)$. In the Appendix we explain in detail how the initial value of $p_y$ is obtained.

The escape of a test particle (star) can easily be determined by using a simple geometrical escape criterion. In particular, an orbit is considered to escape from the scattering region when the star passes over one of the Lagrange points $L_2$ or $L_3$, with velocity pointing outwards.

In previous related works (e.g., \cite{Z15,ZJ17}) we detected a substantial amount of trapped chaotic orbits, for values of the energy just above the energy of escape. These trapped chaotic orbits eventually do escape but only after an extremely long time of numerical integration. On this basis, we decide to classify initial conditions of bounded orbits into two types: (i) non-escaping regular orbits and (ii) trapped chaotic orbits.

Over the years several dynamical indicators have been developed for distinguishing between order and chaos. As in all previous works, we choose to use the Smaller ALingment Index (SALI) method \cite{S01,SM16} which has been proved a very fast yet reliable tool. The mathematical definition of SALI is the following
\begin{equation}
\rm SALI(t) \equiv min(d_-, d_+),
\label{sali}
\end{equation}
where
\begin{align}
d_- &\equiv \norm{ \frac{{\vec{w_1}}(t)}{\| {\vec{w_1}}(t) \|} - \frac{{\vec{w_2}}(t)}{\| {\vec{w_2}}(t) \|} }, \nonumber\\
d_+ &\equiv \norm{ \frac{{\vec{w_1}}(t)}{\| {\vec{w_1}}(t) \|} + \frac{{\vec{w_2}}(t)}{\| {\vec{w_2}}(t) \|} },
\label{align}
\end{align}
are the alignments indices, while ${\vec{w_1}}(t)$ and ${\vec{w_2}}(t)$, are two deviation vectors which initially are orthonormal and point in two random directions. For distinguishing between ordered and chaotic motion, all we have to do is to compute the SALI along a time interval $t_{\rm max}$ of numerical integration. In particular, we track simultaneously the time-evolution of the main orbit itself as well as the two deviation vectors ${\vec{w_1}}(t)$ and ${\vec{w_2}}(t)$ in order to compute the SALI.

The nature of an orbit can be determined by the numerical value of SALI at the end of the numerical integration. Being more precise, if SALI $> 10^{-4}$ the orbit is regular, while if SALI $< 10^{-8}$ the orbit is chaotic (e.g., \cite{SABV04}). On the other hand, when the value of SALI lies in the interval $[10^{-8}, 10^{-4}]$ we have the case of a sticky orbit\footnote{With the term ``sticky orbit" we refer to a special type of orbit which behave as a regular one for long integration time before it exhibits its true chaotic nature.} and further numerical integration is needed to fully reveal the true character of the orbit.

For the numerical integration of the initial conditions of the orbits we set a maximum time of $10^4$ time units. Our previous experience in this system \cite{Z16} suggests that in most cases (energy levels) the majority of the orbits require considerable less time for finding one of the exit channels in the equipotential surface and therefore escape (obviously, the numerical integration is effectively stopped when an orbit escapes through the Lagrange points). Nevertheless, just for being sure that all orbits have enough time to escape, we decided to use such a long integration time. Thus, any orbit which remains bounded inside the Lagrange radius after an integration time of $10^4$ time units will be considered as a bounded (regular or chaotic) one.

Our investigation will be focused on the energy interval $E \in [E(L_2),-0.02]$. We restrict our numerical analysis to this particular energy interval because if the energy is higher than -0.02 then the saddles, over the Lagrange points $L_2$ and $L_3$, are no longer important for the escape of stars.

\begin{figure*}[!t]
\centering
\resizebox{\hsize}{!}{\includegraphics{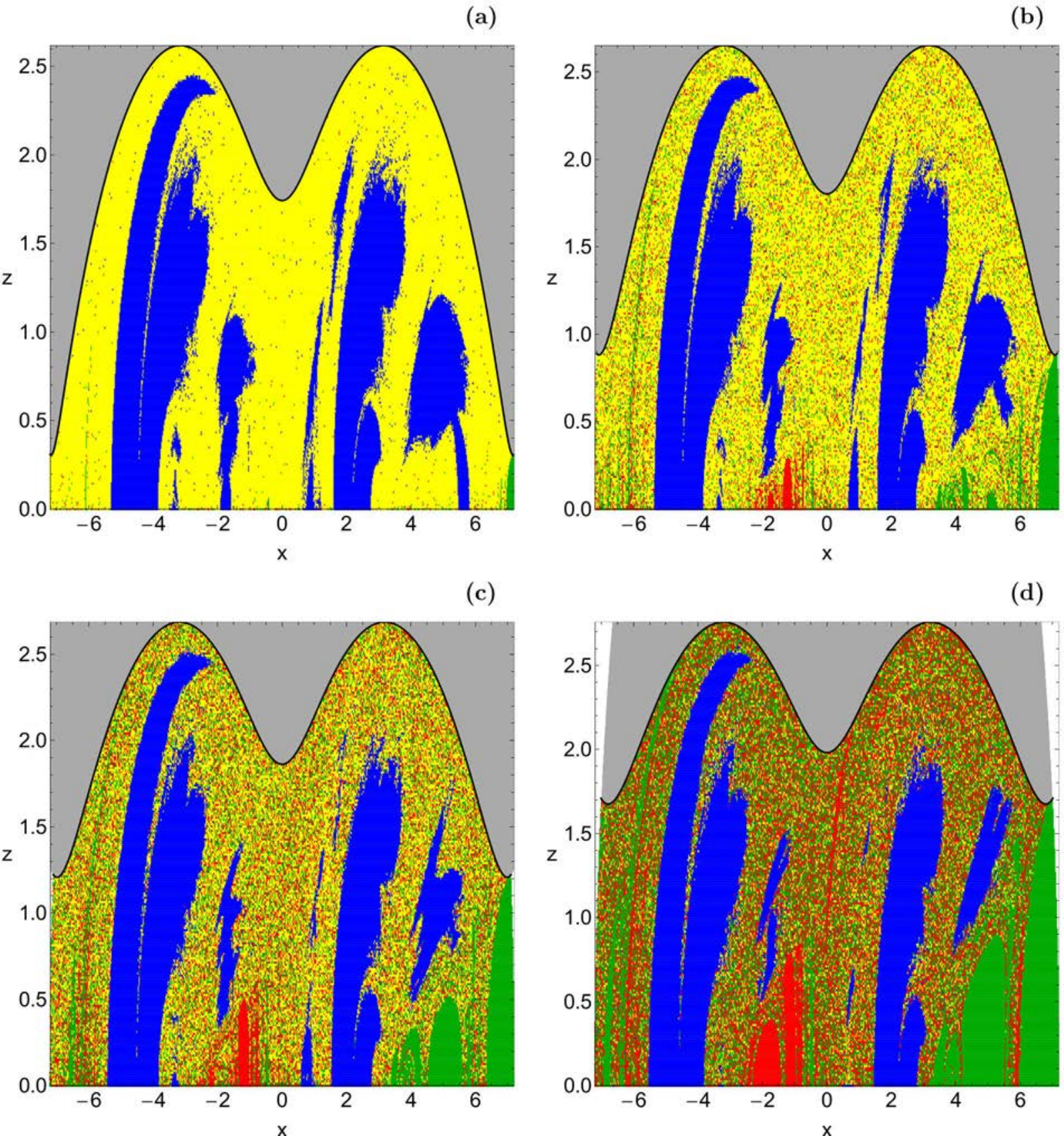}}
\caption{Orbital structure of the $(x,z)$ plane when (a-upper left): $E = -0.0575$; (b-upper right): $E = -0.0570$; (c-lower left): $E = -0.0565$; (d-lower right): $E = -0.0555$. The colour code is as follows: non-escaping regular orbits (blue), trapped chaotic orbits (yellow), sticky orbits (magenta), escaping orbits through $L_2$ (green), escaping orbits through $L_3$ (red). Gray colour corresponds to the energetically forbidden regions. (For the interpretation of references to colour in this figure caption and the corresponding text, the reader is referred to the electronic version of the article.)}
\label{xz1}
\end{figure*}

\begin{figure*}[!t]
\centering
\resizebox{\hsize}{!}{\includegraphics{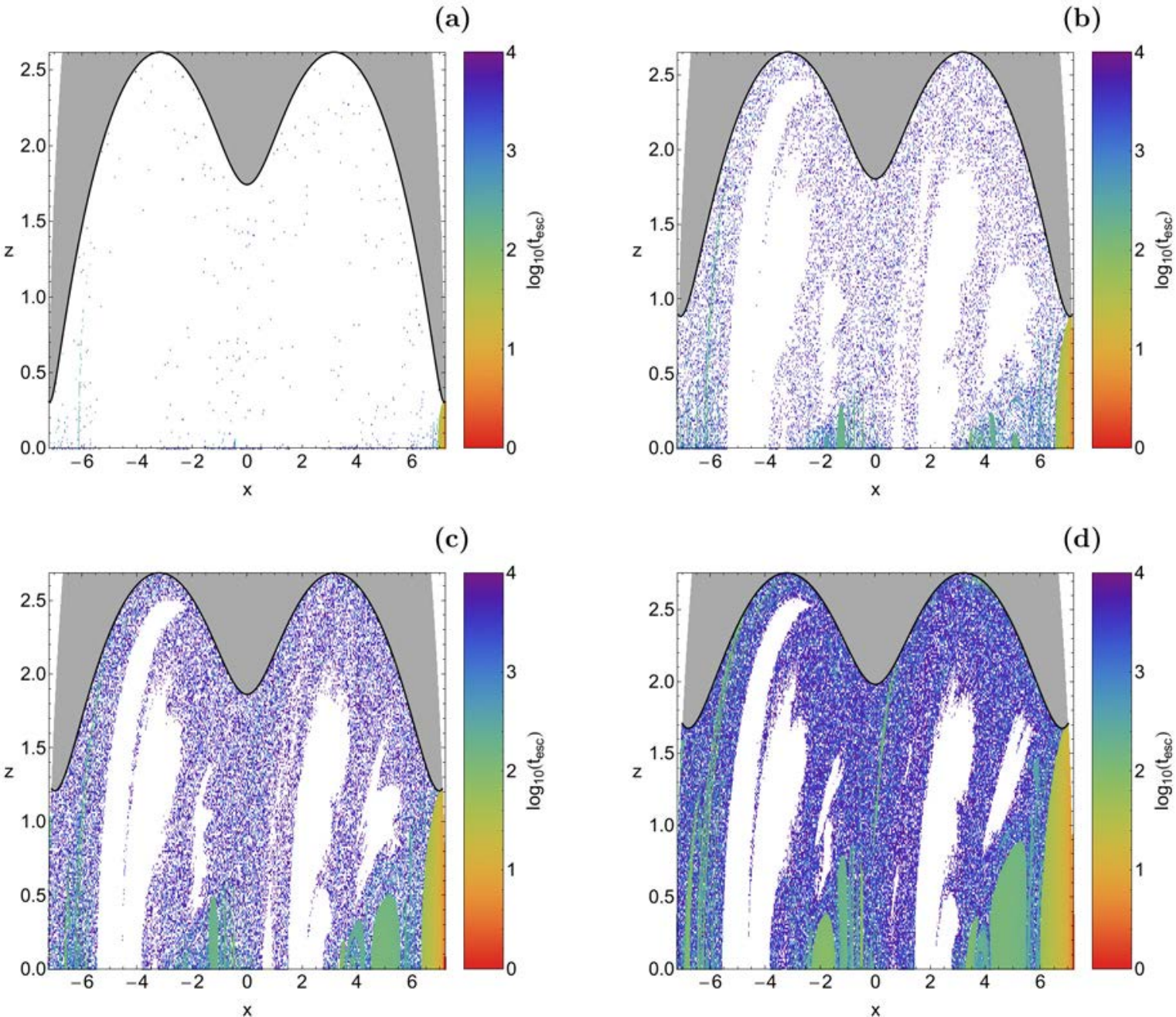}}
\caption{Distribution of the corresponding escape time $t_{\rm esc}$ of the orbits on the $(x,z)$ plane for the values of the energy presented in panels (a-d) of Fig. \ref{xz1}, respectively. The darker the colour, the higher the escape time. Initial conditions of non-escaping regular orbits, sticky orbits and trapped chaotic orbits are shown in white. (For the interpretation of references to colour in this figure caption and the corresponding text, the reader is referred to the electronic version of the article.)}
\label{xz1t}
\end{figure*}

\begin{figure*}[!t]
\centering
\resizebox{0.70\hsize}{!}{\includegraphics{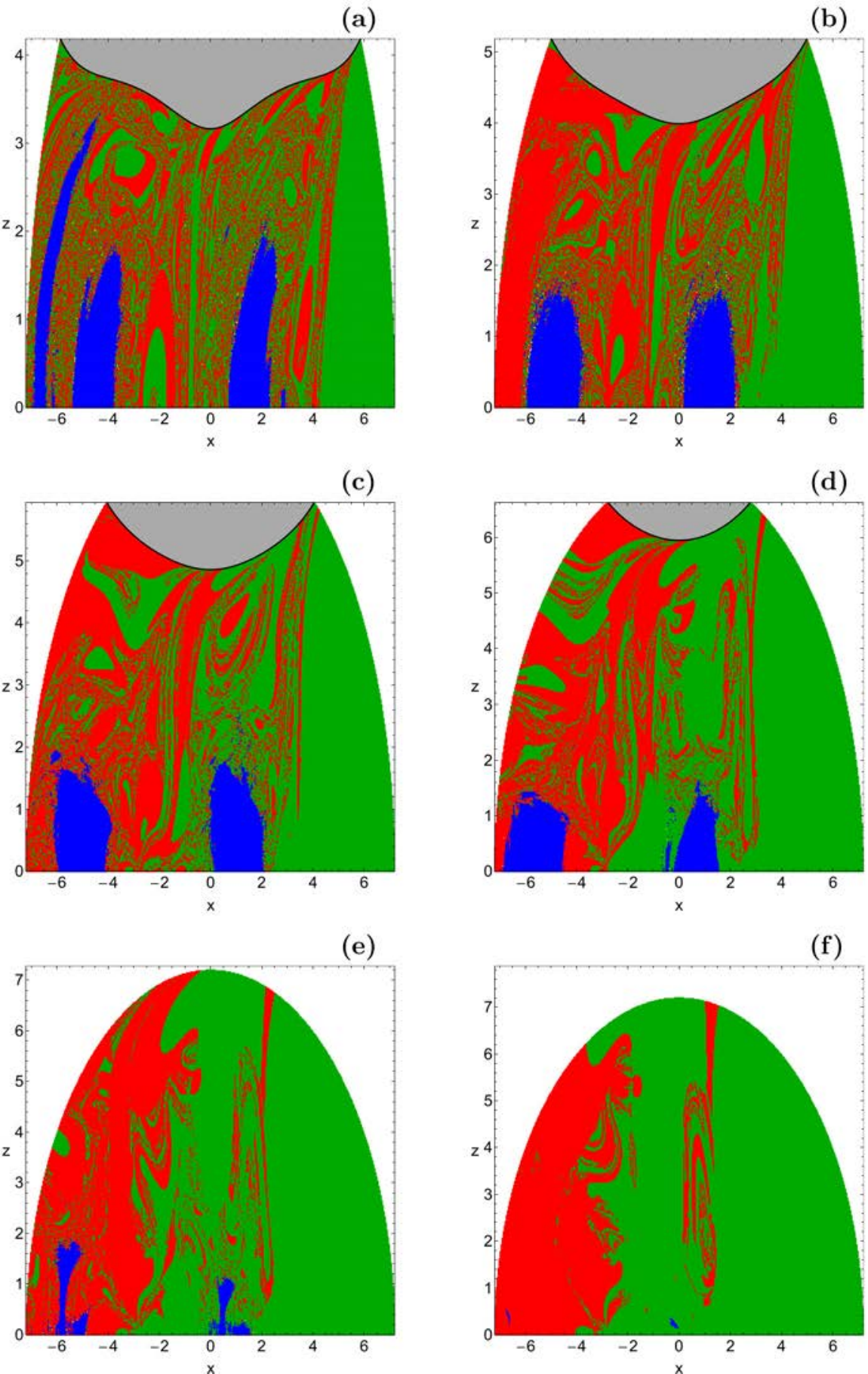}}
\caption{Orbital structure of the $(x,z)$ plane when (a): $E = E(L_4)$; (b): $E = -0.040$; (c): $E = -0.035$; (d): $E = -0.030$; (e): $E = -0.025$; (f): $E = -0.020$. The colour code is the same as in Fig. \ref{xz1}. (For the interpretation of references to colour in this figure caption and the corresponding text, the reader is referred to the electronic version of the article.)}
\label{xz2}
\end{figure*}

\begin{figure*}[!t]
\centering
\resizebox{0.80\hsize}{!}{\includegraphics{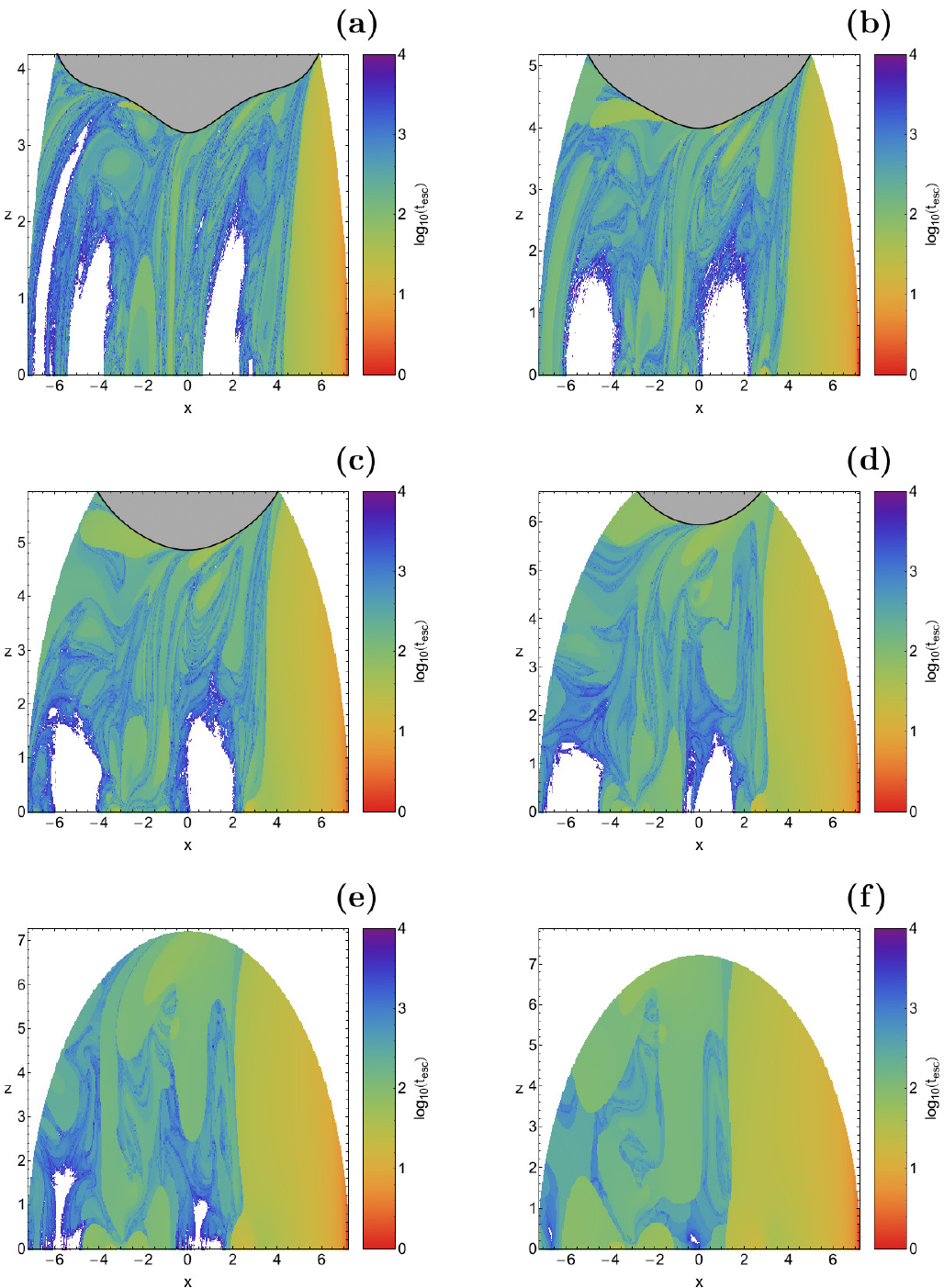}}
\caption{Distribution of the corresponding escape time $t_{\rm esc}$ of the orbits on the $(x,z)$ plane for the values of the energy presented in panels (a-f) of Fig. \ref{xz2}, respectively. The colour code is the same as in Fig. \ref{xz1t}. (For the interpretation of references to colour in this figure caption and the corresponding text, the reader is referred to the electronic version of the article.)}
\label{xz2t}
\end{figure*}

The orbital structure of the $(x,z)$ plane, for four values of the total orbital energy, is presented in Fig. \ref{xz1}(a-d). We present only the $z > 0$ part of the $(x,z)$ plane simply because the $z < 0$ is completely symmetrical with respect to the $x$-axis. In these colour-coded grids we assign to each pixel a specific colour according to the corresponding character of the orbit. In particular, blue colour corresponds to regular non-escaping orbits, yellow colour corresponds to trapped chaotic orbits, magenta colour corresponds to sticky orbits, green colour corresponds to orbits escaping through $L_2$, while the initial conditions of orbits that escape through $L_3$ are marked with red colour. The outermost black solid line denotes the zero velocity curve which is defined as
\begin{equation}
f_1(x,z) = \Phi_{\rm eff}(x, y = 0, z) = E,
\label{zvc1}
\end{equation}
while the energetically forbidden regions are coloured in gray.

In panel (a) of Fig. \ref{xz1} it is seen that for $E = -0.0575$, which is an energy level just above the energy of escape, more than two thirds of the $(x,z)$ plane are occupied by initial conditions which correspond to trapped chaotic orbits. However inside the vast trapped chaotic sea we can locate isolated initial conditions of escaping orbits. Nevertheless, it is evident that for energy levels above yet such close to the critical energy $E(L_2)$ the phenomenon of escaping stars is extremely rare and very difficult to occur. As we move away from the critical energy level $E(L_2)$ we observe in panels (b-d) of Fig. \ref{xz1} that the trapped chaotic sea weakens, while at the same time the amount of escaping orbits increases rapidly. Moreover local basins of escape emerge mainly either in the area between the two galaxies or near the vicinity of the Lagrange point $L_2$. At this point, we would like to note that by the term ``basin of escape" we refer to a local set of initial conditions of orbits that corresponds to a certain escape channel. For $E = -0.0555$ we see in panel (d) of Fig. \ref{xz1} that the majority of the $(x,z)$ plane is covered by a highly fractal\footnote{It should be emphasized that when we state that an area is ``fractal" we simply mean that it has a fractal-like geometry without any further calculation of the fractal dimension, as in \cite{AVS01,AVS09}.} mixture of escaping and trapped chaotic orbits. Inside this fractal domain the escape mechanism of the orbits exhibits a high dependence on the particular initial conditions. This phenomenon strongly implies that a minor change in the $(x_0,z_0)$ initial conditions of the orbit has as a result the star to escape through the opposite escape channel, which is of course a classical indication of chaotic motion.

The corresponding distribution of the escape time $t_{\rm esc}$ of the orbits presented in Fig. \ref{xz1}(a-d) is illustrated in Fig. \ref{xz1t}(a-d), where the scale of the escape time is revealed through a colour code. Specifically, initial conditions of fast escaping orbits are indicated with light reddish colors, while dark blue/purple colors suggest high escape periods. Furthermore, initial conditions of non-escaping, sticky, and trapped chaotic obits are shown in white. Note that the colour bar has a logarithmic scale. It is observed in panels (a-c) that the escape time of the orbits are huge corresponding to several thousands of time units. However, as the value of the energy increases the escape rates of the orbits are constantly being reduced. Looking carefully at the distributions of the escape time it becomes more than evident that orbits with initial conditions in the fractal areas of the $(x,z)$ planes require a significant amount of time before they escape from the binary system. On the other hand, the shortest escape rates have been measured for orbits inside the basins of escape, where there is no dependence on the initial conditions whatsoever (see e.g., panel (d) of Fig. \ref{xz1t}).

Our numerical calculations indicate that for $E > -0.0535$ trapped chaotic motion is almost negligible as the corresponding initial conditions appear only as lonely, isolated points either near the vicinity of the boundaries of the stability islands or randomly scattered in the vast escape region.

\begin{figure}[!t]
\begin{center}
\includegraphics[width=\hsize]{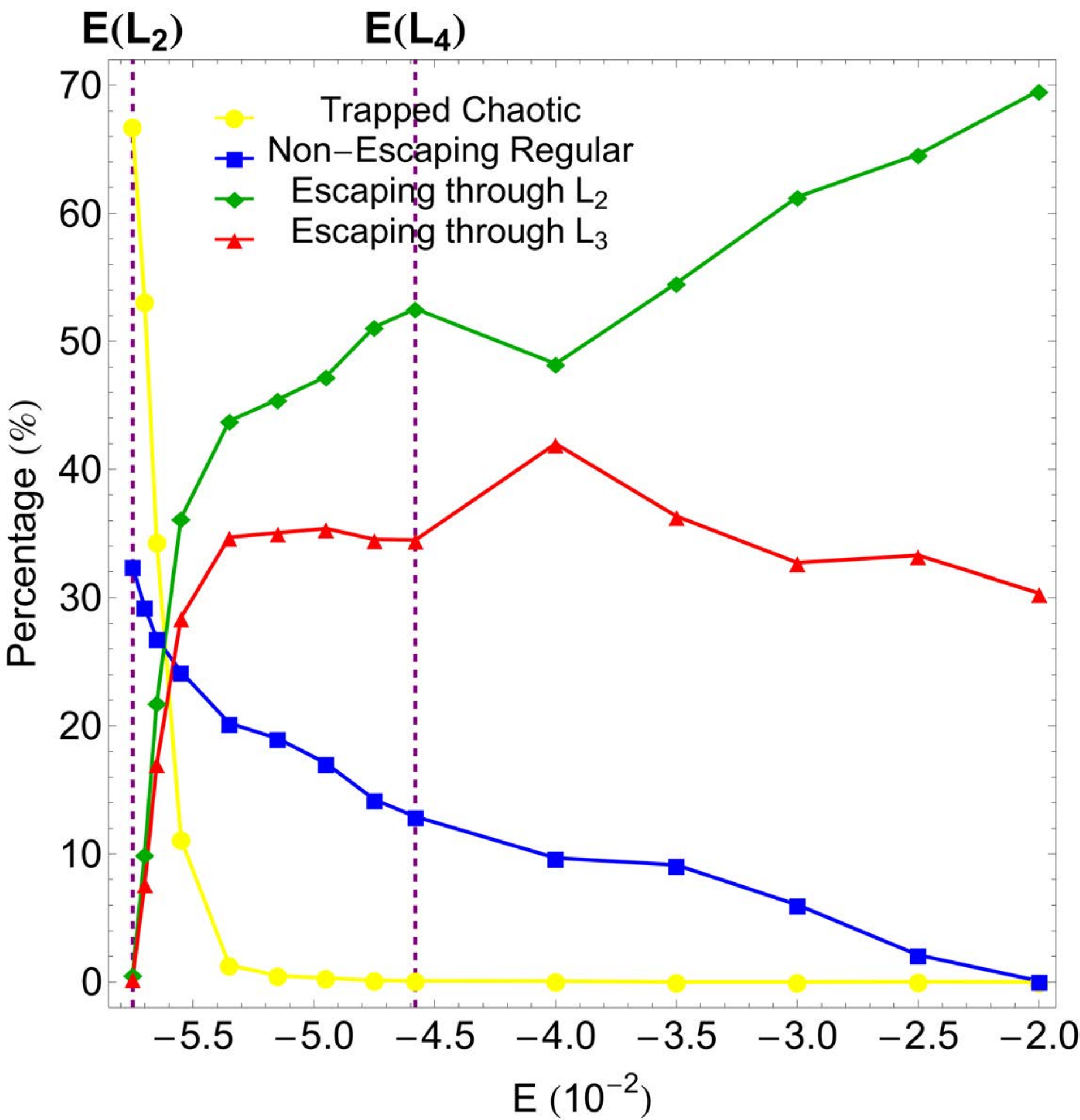}
\end{center}
\caption{Evolution of the percentages of all types of orbits on the $(x,z)$ plane, as a function of the total orbital energy $E$. The vertical, dashed, magenta lines indicate the critical energy levels $E(L_2)$ and $E(L_4)$. The distribution shown applies to a particular plane of initial conditions. The asymmetry between the escape through $L_2$ and $L_3$ is caused by the particular choice of one branch of $p_{y0}$ (i.e. one orientation of the initial intersection of the plane $y = 0$). (For the interpretation of references to colour in this figure caption and the corresponding text, the reader is referred to the electronic version of the article.)}
\label{percs}
\end{figure}

\begin{figure}[!t]
\begin{center}
\includegraphics[width=\hsize]{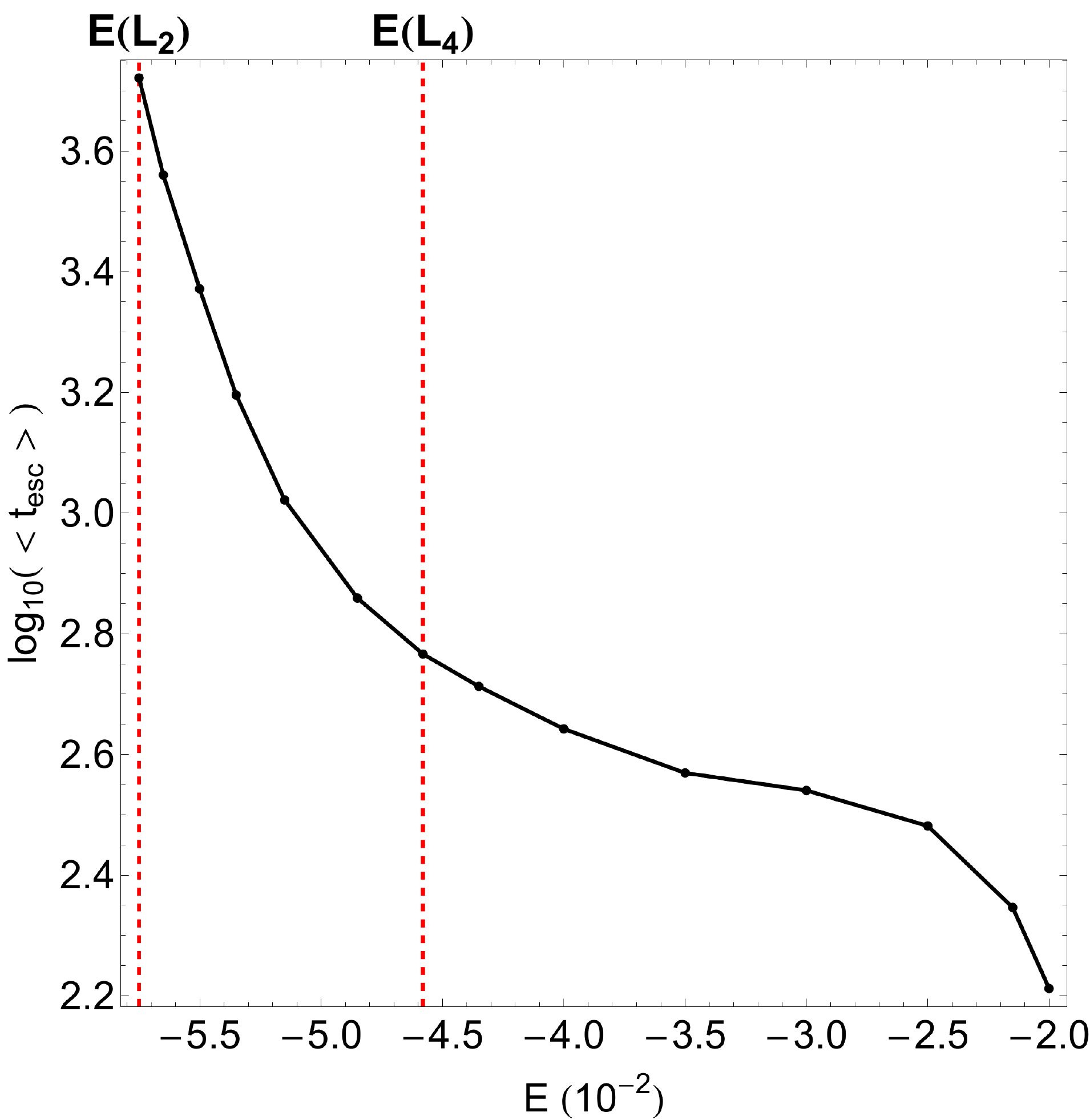}
\end{center}
\caption{Evolution of the logarithm of the average escape time of the orbits $(\log_{10}\left( < t_{\rm esc} > \right) )$, as a function of the total orbital energy $E$. The vertical, dashed, red lines indicate the critical energy levels $E(L_2)$ and $E(L_4)$. (For the interpretation of references to colour in this figure caption and the corresponding text, the reader is referred to the electronic version of the article.)}
\label{tavg}
\end{figure}

The escape dynamics of the binary system for six additional values of the total orbital energy is given in Fig. \ref{xz2}(a-f). Once more, different colours are used for distinguishing between the five types of the orbits (non-escaping regular, sticky, trapped chaotic, escaping through $L_2$ and escaping through $L_3$). In this case all six energy levels are much higher than the energy of escape. In panel (a) of Fig. \ref{xz2}, where $E = E(L_4)$, we observe that about half of the energetically allowed area, inside the Lagrange radius, is occupied by initial conditions of escaping orbits, which form well-defines basins of escape. As the value of the total orbital energy increases two phenomena take place: (i) the amount of the escaping orbits and the corresponding basins of escape grow significantly and (ii) the area of the non-escaping regular orbits is reduced. When $E = -0.025$ it is seen in panel (e) of Fig. \ref{xz2} that the stability islands around both galaxies are very confined, while in panel (f) of the same figure, for $E = -0.020$ the stability islands are hardly visible, while the basins of escape completely dominate the $(x,z)$ plane. In particular, basins of escape corresponding to escape through $L_2$ cover about 70\% of the energetically allowed area.

In Fig. \ref{xz2t} we illustrate the corresponding distribution of the escape time of the orbits, for the values of the energy presented in panels (a-f) of Fig. \ref{xz2}, respectively. One may observe that in general terms the escape time of the orbits for these energy levels is significantly lower, with respect to what we seen earlier in Fig. \ref{xz1t}. However, even for such high energy levels $(E > E(L_4))$ we still have a substantial portion of orbits with relatively high escape time. These orbits have mainly initial conditions either in the fractal escape basin boundaries, or near the boundaries of the stability islands.

When $E(L_2) < E < E(L_4)$ a significant portion of the $(x,z)$ planes corresponds to stability islands of non-escaping regular orbits (see e.g., Fig. \ref{xz1}(a-d)). Additional numerical computations reveal that almost all the regular orbits are in fact 1:1:0\footnote{The $n:m:l$ notation we use for classifying the regular three-dimensional (3D) orbits is according to \cite{CA98}. The ratio of those integers corresponds to the ratio of the main frequencies of the orbit, where the main frequency is the frequency of the greatest amplitude for each coordinate. Main amplitudes, when having a rational ratio, define the resonance of an orbit.} resonant loop orbits, which circulate parallel to the $(x,y)$ plane. Furthermore, our spectral analysis indicates that there are several types of loop orbits around each galaxy. However we found that the only possible regular motion of stars is to circulate only around one of the two galaxies. On the contrary, there is no numerical indication of stars moving in regular orbits around both galaxies.

There is doubt that the evolution of the percentages of all types of orbits as a function of the total orbital energy $E$ will add considerably to our knowledge regarding the escape dynamics of the binary galaxy system. In Fig. \ref{percs} we present the corresponding diagram. Here it should be emphasized that for constructing this diagram we used numerical data from additional colour-coded grids, apart from those shown in Fig. \ref{xz1} and Fig. \ref{xz2}. We see that just above the critical energy level $E(L_2)$ trapped chaotic orbits dominate by occupying more than 65\% of the energetically allowed area on the $(x,z)$ plane. However, as we move away from the energy of escape the rate of trapped chaotic orbits is heavily being reduced, while for $E > -0.05$ it completely vanishes. The percentage of non-escaping regular orbits starts at about 32\%, for $E = -0.0575$, and it is constantly being reduced until $E = -0.02$, where it is zeroed. The parametric evolution of the rates of the escaping orbits on the other hand is completely different, with respect to those of bounded (regular and chaotic) orbits. In particular, we observe that for $E(L_2) < E < -0.0565$ the rate of escaping orbits, through $L_2$ and $L_3$, are very close, while for $E > -0.0565$ they start to diverge. Being more precise, the amount of escaping orbits through $L_2$ increases more rapidly, with respect to the portion of escaping orbits through $L_3$. For $E = -0.04$ both types of escaping orbits display a sudden change (peak or drop), for which we have no explanation yet. For $E > -0.04$ the percentage of escaping orbits through $L_2$ continues to increase up to 70\%, while that of escaping orbits through $L_3$ is being reduced until $E > -0.03$ where it seems to saturate around 30\%. Therefore taking into consideration all the above-mentioned analysis one may reasonably deduce that at low energy levels, where the fractality of the $(x,z)$ plane is maximum, stars do not show any particular preference regarding the escape channels and bounded motion (regular and chaotic) is the most populated type of motion. On the contrary, at high enough energy levels $(E \to -0.02)$, where basins of escape completely dominate, it seems that escape through $L_2$ is twice more preferable, with respect to escape through $L_3$.

\begin{figure*}[!t]
\centering
\resizebox{0.80\hsize}{!}{\includegraphics{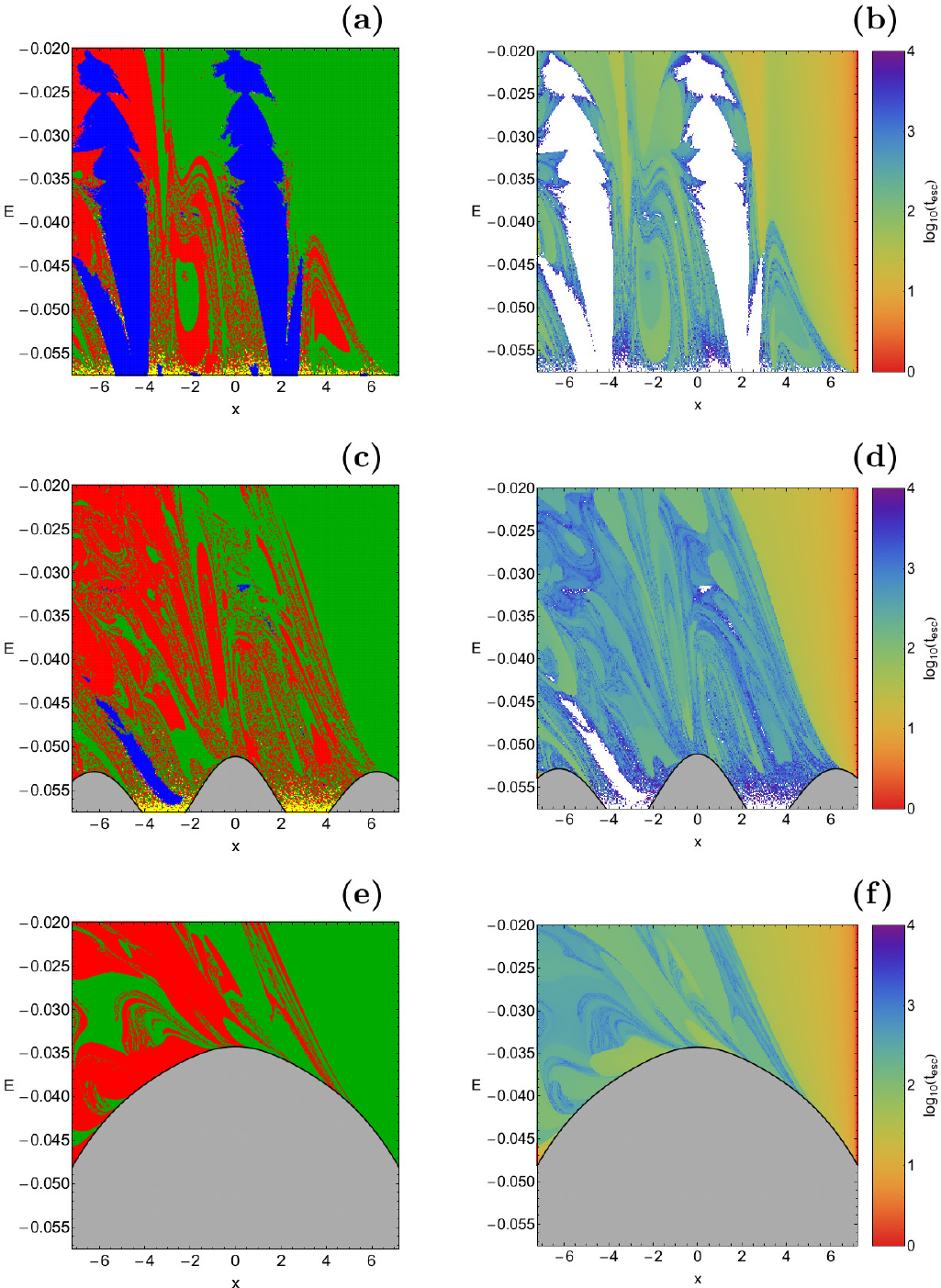}}
\caption{(left column): Orbital structure of the $(x,E)$ plane when (a): $z_0 = 0.01$; (b): $z_0 = 2.5$; (c): $z_0 = 5.0$. The colour code is the same as in Fig. \ref{xz1}. (right column): Distribution of the corresponding escape time $t_{\rm esc}$ of the orbits on the $(x,E)$ plane for the corresponding values of $z_0$. (For the interpretation of references to colour in this figure caption and the corresponding text, the reader is referred to the electronic version of the article.)}
\label{xEt}
\end{figure*}

Needless to say that the relative fraction of orbits escaping through saddles $L_2$ and $L_3$ strongly depends on our particular choice of the initial conditions of the orbits. The inversion symmetry of the whole system mentioned before implies that to each orbit escaping through $L_2$ there is another orbit, with inverted initial conditions, which escapes through the saddle $L_3$. Therefore an integration over the whole 5-dimensional energy shell of possible initial conditions would give equal escape rates through both escape channels. And also a random collection of initial conditions from the
interior part of the 5-dimensional energy shell would lead to equal escape rates. On the other hand, for grids of initial conditions on any particular lower dimensional surface, such as those on the $(x,z)$ plane, these escape rates depend on how this particular surface intersects the basins of escape belonging to the two saddles.

At this point we would like to point out that our numerical computations suggest that sticky orbits always possess an extremely low percentage (less than 0.001\%), which strongly implies that the chosen total time of the numerical integration $(t_{\rm max} = 10^4)$ is indeed sufficient, so as almost all sticky orbits to be correctly classified as chaotic ones.

The evolution of the average value of the escape time $< t_{\rm esc} >$ of the orbits as a function of the total orbital energy $E$ is given in Fig. \ref{tavg}. One may observe that very close to $E(L_2)$ the average escape time of the orbits corresponds to more than 4000 time units. With increasing energy however, the required time for escape is constantly being reduced until $E = -0.025$, where $< t_{\rm esc} > \simeq 320$ time units. For higher values of the energy the reduction of $t_{\rm esc}$ continues but with a different slope (more rapidly). It is interesting to note that the parametric evolution of the average value of the escape time of the orbits is extremely smooth, without displaying any types of abnormalities.

The colour-coded diagrams presented in Figs. \ref{xz1} and \ref{xz2} can provide useful information, regarding the orbital structure of the $(x,z)$ plane, however only for some specific values of the total orbital energy $E$. In order to overcome this limitation and obtain a more complete view of the orbital structure of the dynamical system, we shall adopt the H\'{e}non's method \cite{H69}, thus trying to gather information of a continuous spectrum of energy values. In particular, for specific values of $z_0$ we define dense uniform grids of initial conditions on the $(x,E)$ plane with $y_0 = p_{x_0} = p_{z_0} = 0$, while in all cases the initial value of $p_y$ is obtained from the Hamiltonian (\ref{ham}). Using this method we are able to monitor the evolution of the orbital structure of the system, using a continuous spectrum of energy values, rather than a few discrete ones. The orbital structure of the $(x,E)$ plane, when $E \in [E(L_2),-0.02]$, along with the distribution of the corresponding escape time of the orbits, is presented in Fig. \ref{xEt}(a-f). The black solid line is the limiting curve which in this case is defined as
\begin{equation}
f_2(x,E;z_0) = \Phi_{\rm eff}(x,y = 0,z = z_0) = E.
\label{zvc2}
\end{equation}

Panels (a) and (b) of Fig. \ref{xEt} correspond to $z_0 = 0.01$, that is a low value of initial $z$. In other words this means that the three dimensional orbits are started very close to the $(x,y)$ plane. We observe that throughout the energy range stability islands of 1:1:0 loop orbits are present near the centres of both galaxies, while the rest of the $(x,E)$ plane is covered either by well-formed basins of escape or by highly fractal domains. However for $E > -0.03$ the fractal regions disappear and basins of escape dominate. When $z_0 = 2.5$ we see in panels (c) and (d) of Fig. \ref{xEt} that bounded basins of non-escaping regular motion are very limited and appear mainly for relatively low values of the energy $(E < -0.04)$. Moreover, the majority of the $(x,E)$ plane displays a very complicated structure with multiple basins of escape, surrounded by highly fractal regions. Finally, for $z_0 = 5$ (see panels (e) and (f) of Fig. \ref{xEt}) there is no numerical indication of ordered bounded motion and all the examined initial conditions of orbits escape. It is interesting to note that as the initial value of the $z$ coordinate increases the energetically allowed regions of motion appear for higher energy levels, while at the same time the area of the energetically forbidden region grows rapidly.

For the numerical integration of the initial conditions of the three-dimensional orbits, in all types of colour-coded grid presented in this Section, we needed roughly about 2.5 days of CPU time on an Intel$^{\circledR}$ Quad-Core\textsuperscript{TM} i7 2.4 GHz PC.

\section{The development scenario of the saddle dynamics}
\label{nhims}

\subsection{Invariant subsets over potential saddles}
\label{ss1}

When a dynamical system is given and we want to understand the development scenario of the dynamics as a function of the energy, or of some other parameter, then we first identify and investigate the most important invariant subsets of the phase space. Certainly this task includes the study of the most fundamental periodic orbits. They belong to the central elements in the skeleton of the dynamics. When we have a system of 3 or more degrees of freedom, then also subsets of higher dimension play an important role and must be understood well. In most cases it is easier and more instructive to investigate the dynamics in an appropriate Poincar\'e map for a fixed energy than to investigate the flow. Also we will work to a large extent with the maps.

First one has to understand how and under what circumstances an invariant subset can be important for the global dynamics. Some clue is given by the experience with the Poincar\'e maps for 2-dof systems. These maps act on a 2-dimensional domain.
Their most important invariant sets are fixed points and in particular the hyperbolic ones. These fixed points have 1-dimensional stable and unstable manifolds which divide the domain into pieces of different behaviour. Usually we find such fixed points (and their corresponding unstable periodic orbits in the flow) over saddle points of the effective potential of the system. The important feature of these saddle fixed points/saddle orbits is that their stable and unstable manifolds direct the flow over the saddle and thereby direct and channel the escape from inner potential holes to the outer parts of the system. We obtain the corresponding escape properties in position space by a projection from phase space to position space. Generally speaking, orbits approach the saddle region along the stable manifolds of the invariant set over the saddle and leave the saddle region along their unstable manifolds. Thereby it is already evident that the unstable manifolds trace out the parts of space through which the escaping orbits run and they define the spatial patterns created by the escaping orbits.

For 2-dof systems these ideas are easy to understand and to visualize. How are these ideas generalized for more degrees of
freedom and their higher dimensional phase spaces and higher dimensional domains of the Poincar\'e maps ? To get an idea we first have to identify the properties of the invariant sets which cause their global importance for the dynamics and then we have to conserve these important properties under the increase of dimensions. Essential for the property to channel and direct the flow over the saddle is that the stable and unstable manifolds of the invariant objects are of codimension 1, this holds as well in the flow as in the map. This value of the codimension guarantees that these manifolds separate distinct sides (at least locally) and it gives them the possibility to cut the phase space/domain of the map into bits and pieces of qualitatively different behaviour.

This picture suggests that also in higher dimensional systems we should be looking for invariant subsets having stable and unstable manifolds of codimension 1. Then the invariant subsets of interest themselves must be of codimension 2 in the domain of the map or correspondingly also in the flow. And of course in order to produce stable and unstable manifolds the invariant subsets must be hyperbolic (unstable) in the 2 directions transverse to them and this transverse instability must dominate
any possible tangential instability. The mathematics of such invariant subsets are known for a long time, they are the normally hyperbolic invariant manifolds (NHIMs) of codimension 2. These are the subsets in the phase space and in the map we are looking for in this section.

Now a natural question arises: Where do we usually find such objects? The answer is: over index-1 saddles of the effective potentials. This becomes clear as follows: For energies very close, but above the saddle energy, take the quadratic expansion of the effective potential around the saddle point. The corresponding approximate equations of motion are linear. For a system with $n$ degrees of freedom there are $n$ normal modes over the saddle. For an index-1 saddle $n-1$ of them are stable, 1 is unstable. Accordingly a general orbit starting over the saddle remains in the saddle region for all times forward and backward, if it is composed of the stable modes only and does not contain any contribution from the unstable
mode. There is a $(2n-4)$-dimensional continuum of such orbits. Namely we have a $(n-2)$-dimensional continuum of choices for the distribution of the available energy of motion between the $n-1$ stable modes and we have a $(n-2)$-dimensional continuum of choices for the relative phases between the contributions of the $n-1$ stable modes. This $(2n-4)$-dimensional continuum of orbits forms a $(2n-3)$-dimensional surface in the energy shell of the phase space. In the quadratic approximation this surface is an ellipsoid. In the Poincar\'e map we obtain a corresponding $(2n-4)$-dimensional invariant surface in its domain. These are the sought for saddle NHIMs of codimension 2 in quadratic approximation. All orbits in such NHIMs are unstable under the smallest perturbation in the direction of the unstable mode. Accordingly the whole invariant surface is dynamically unstable in normal directions.

What happens under the inclusion of the anharmonicities in the effective potential, which become important if we increase the energy? Or more generally: What happens to this invariant subset under general perturbations? Here the persistence property of NHIMs becomes important. NHIMs survive perturbations as long as they are dominated by the transverse instability, i.e. as long as the transverse instability is larger than any tangential instability. As long as the normal hyperbolicity is conserved, the NHIM surface may become smoothly deformed under perturbations, it may be shifted a little, but it remains a smooth surface with the topology of a sphere. Only for large perturbations the normal hyperbolicity may be lost and then the surface may be destroyed or may be transformed into something completely different.

Our system is a 3-dof system, therefore in the following we are interested in the case $n=3$. In \cite{JZ16} we have given all the equations for the Lyapunov orbits and for the NHIM orbits in the quadratic approximation. These equations hold for any index-1 saddle of a 3-dof Hamiltonian system. We shall not repeat these equations here and the reader is referred to the above-mentioned reference. Also many explanations in subsections 4.1 and 4.2 of \cite{JZ16} transfer word by word to the present system.

\subsection{Scenario of the fundamental periodic orbits}
\label{ss2}

To the stable modes of motion over the saddle there correspond simple periodic orbits, the Lyapunov orbits, which realize a
single stable mode in pure form. These orbits are the most important orbits within the NHIMs. Therefore a good understanding of the development scenario of the Lyapunov orbits and other periodic orbits related to them helps a lot to understand the development scenario of the whole NHIM and consequently of the whole system. In this subsection we want to understand the development scenario of the NHIMs over the index-1 saddles, under an increase of the total orbital energy $E$. Therefore we start in this subsection with the scenario of the most important periodic orbits in the system.

\begin{figure}[!t]
\includegraphics[width=\hsize]{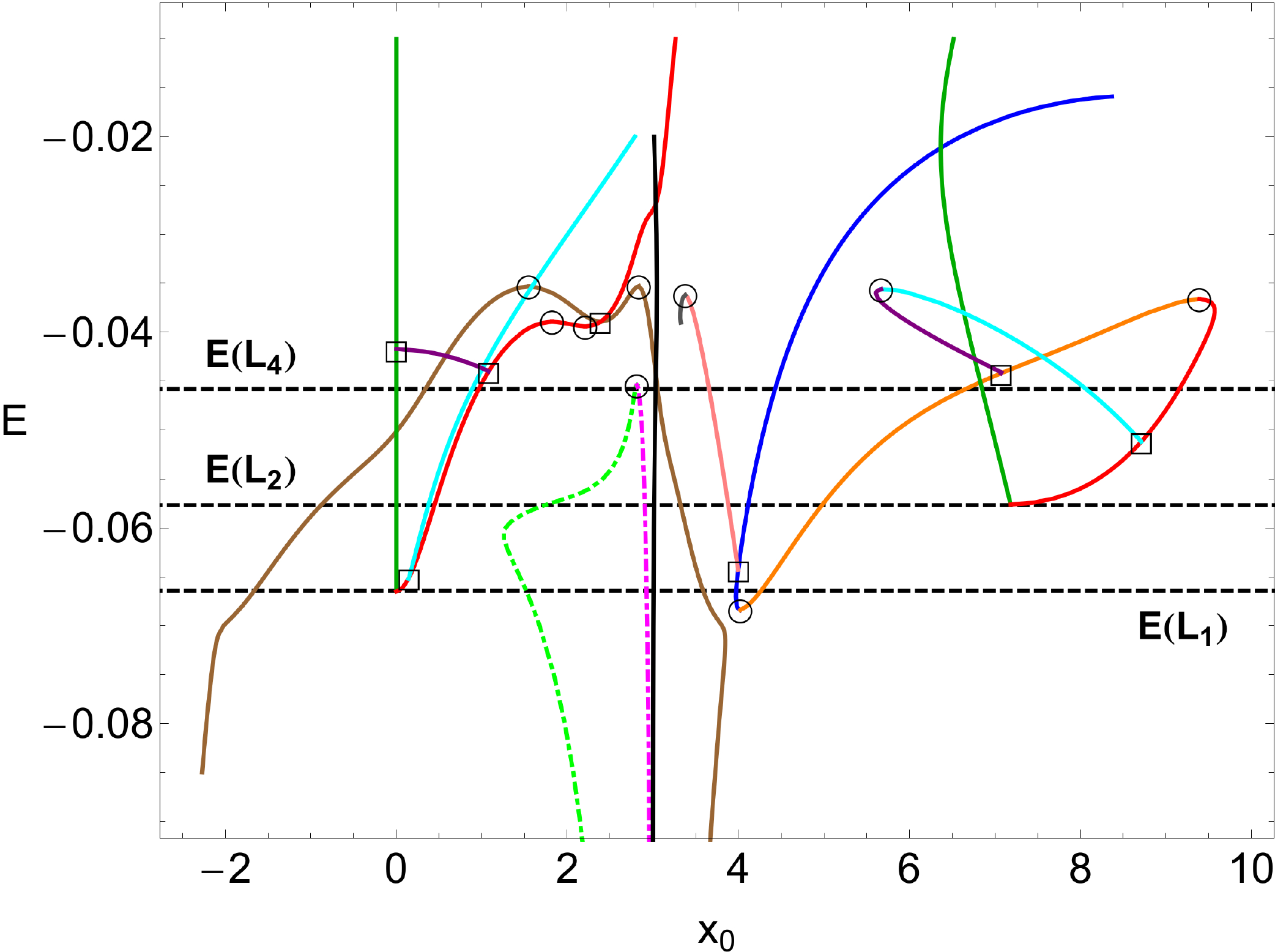}
\caption{Parametric evolution of the $x_0$ initial condition of the most important periodic orbits, as a function of the total orbital energy $E$. The horizontal black dashed lines correspond to the critical levels of the energy. The colour code is explained in the text. (For the interpretation of references to colour in this figure caption and the corresponding text, the reader is referred to the electronic version of the article.)}
\label{fpo}
\end{figure}

In Fig. \ref{fpo} we present a bifurcation diagram where we plot as a function of the energy the value of the $x$-coordinate of each important periodic orbit, at the moment when it crosses the plane $y = 0$ in negative orientation. Dashed lines mark horizontal orbits, i.e. orbits lying completely in the plane $z = 0$, which circle in positive orientation. They appear as light green and magenta lines in the figures. Solid lines on the other hand, mark orbits which circle in the plane $z = 0$ in negative orientation or also orbits moving out of plane. A further distinction of the type of orbits plotted by solid lines is done by the colours. In particular
\begin{itemize}
  \item Red indicates horizontal orbits coming from the index-1 saddles.
  \item Dark green indicates vertical orbits coming from the index-1 saddles.
  \item Brown indicates horizontal orbits coming out of the potential wells.
  \item Black indicates vertical orbits coming out of the potential wells.
  \item Blue, pink, grey and orange are further horizontal orbits.
  \item Cyan and purple are tilted loop orbits.
\end{itemize}
Furthermore, black open circles mark saddle-centre bifurcations of the involved orbits, while black open squares mark the most important pitchfork bifurcations. We mainly show the orbits coming from the potential hole around $P_2$ and from the saddles $L_1$ and $L_2$. In the hole around $P_1$ the scenario from the other hole is repeated symmetrically and over the saddle $L_3$ the scenario from the saddle $L_2$ is repeated. However, note that the position of orbits from the two symmetric sides of the effective potential do not appear symmetrically in the figure since we select the position of the negative intersection with the plane $y = 0$ and this breaks the appearance of the symmetry.

For energies $E(P_1) < E < E(L_1)$ the accessible part of the position space has 3 disjoint components, the outer part and 2 potential wells. The outer parts do not have periodic orbits. Therefore the first periodic orbits in the system are born at $E = E(P_1)$. The point $P_1$ is a minimum of the potential, therefore for energies close to $E(P_1)$ the motion in the potential well has 3 stable modes and 3 corresponding fundamental periodic orbits. There is one orbit oscillating up and down in the $z$-direction, let us call this one mv2 in the following, it is of little importance for the scenario, it is marked by black. And there are two periodic orbits rotating in the plane $z=0$. One of these orbits circles in positive orientation around the point $P_1$, let us call this one mi2, also it will not play any important role in the following, it is marked light green. A further orbit of this type is marked magenta. The other horizontal orbit circulates in negative orientation, let us call it mh2, it is plotted brown, it plays an important role for the whole scenario.

\begin{figure}[!t]
\includegraphics[width=\hsize]{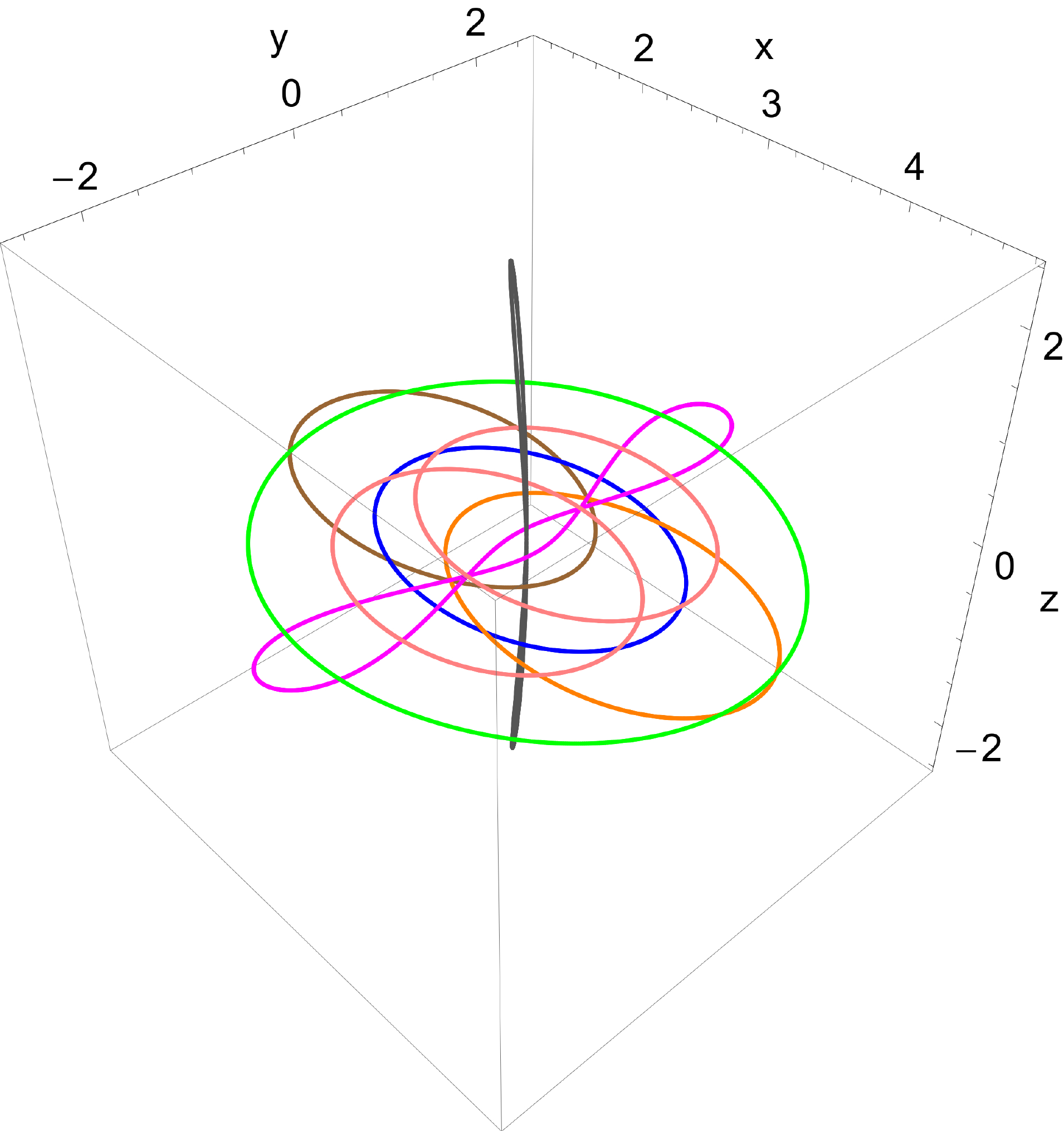}
\caption{Collection of the periodic orbits, in position $(x,y,z)$ space, near the potential well, when $E = -0.06$. The color code is the same as in Fig. \ref{fpo}. (For the interpretation of references to colour in this figure caption and the corresponding text, the reader is referred to the electronic version of the article.)}
\label{orbsp2}
\end{figure}

\begin{figure*}[!t]
\centering
\resizebox{\hsize}{!}{\includegraphics{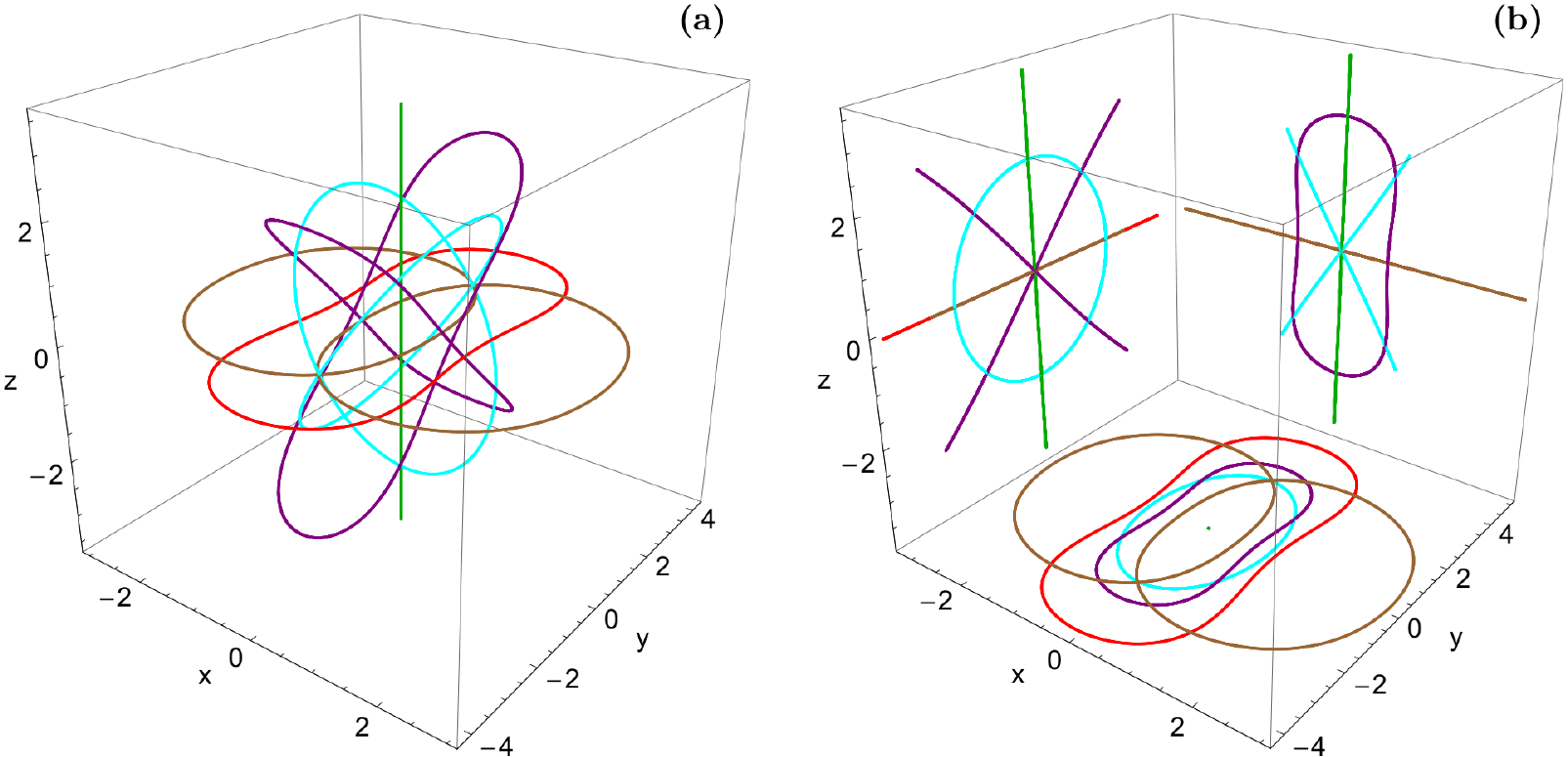}}
\caption{(a-left): A collection of the most important periodic orbits in the configuration $(x,y,z)$ space, near $L_1$, when $E = -0.043$. (b-right): The projections of the periodic orbits into the primary planes $(x,y)$, $(x,z)$ and $(y,z)$. The colour code is the same as in Fig. \ref{fpo}. (For the interpretation of references to colour in this figure caption and the corresponding text, the reader is referred to the electronic version of the article.)}
\label{orbsl1}
\end{figure*}

\begin{figure*}[!t]
\centering
\resizebox{\hsize}{!}{\includegraphics{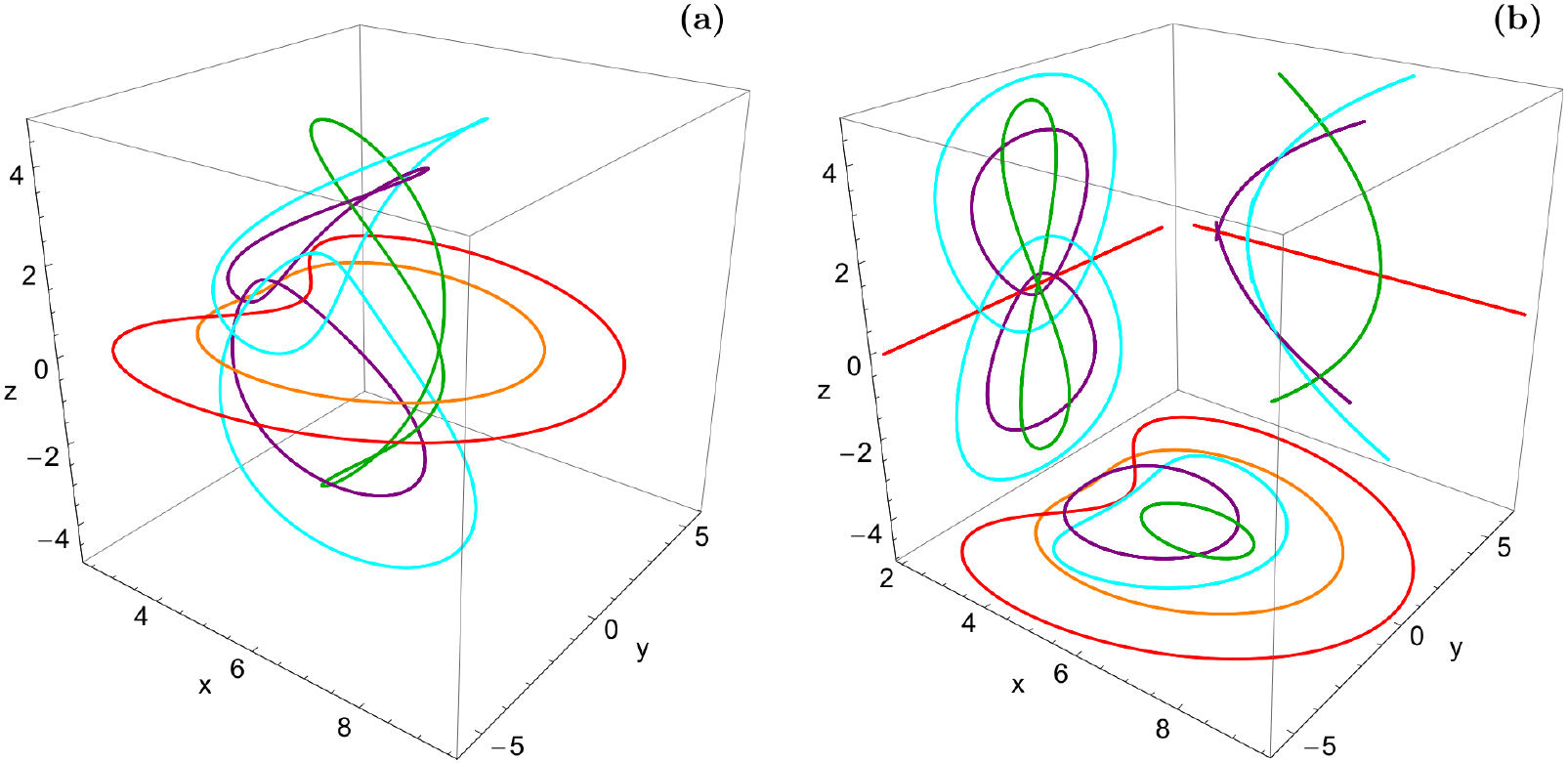}}
\caption{(a-left): A collection of the most important periodic orbits in the configuration $(x,y,z)$ space, near $L_2$, when $E = -0.04$. (b-right): The projections of the periodic orbits into the primary planes $(x,y)$, $(x,z)$ and $(y,z)$. The colour code is the same as in Fig. \ref{fpo}. (For the interpretation of references to colour in this figure caption and the corresponding text, the reader is referred to the electronic version of the article.)}
\label{orbsl2}
\end{figure*}

Around the energy -0.068 there is a saddle-centre bifurcation in the potential wells which creates the new orbits ba2 and bb2, marked blue and orange respectively. We see that the orbits mh2, ba2 and bb2 together form a kind of symmetry broken pitchfork bifurcation (for this phenomenon see also Fig. 7.2 in \cite{LL83} or Fig. 20.3.2 in \cite{W03}). For energies above -0.066 the orbit ba2 acts as a natural continuation of mh2 for small energies and mh2 for high energy and bb2 act as if they would be orbits split off from mh2 for small energy. At $E \approx -0.064$ ba2 suffers a pitchfork bifurcation where it splits off the pair br2a/br2b, marked pink. Because of symmetry reasons this pair creates a single curve only in the diagram. This also holds for many other pitchfork pairs which show up in the diagram. This pair of orbits br2a/br2b disappears at $E \approx -0.0362$ in a pair of saddle-centre bifurcations where it collides with another pair of horizontal orbits marked grey in Fig. \ref{fpo}. These pink and grey orbits are of little importance for the global bifurcation diagram of Fig. \ref{fpo}. However, in subsection \ref{ss6} we will make a comment on the energy value where these orbits disappear. Into the bifurcation plot we have also included the orbit mh1 which is the symmetry counter part to mh2 and it comes from the potential hole around $P_1$. For $E > -0.05$ this orbit has already grown quite large and in addition it has moved in direction to the saddle $L_1$, i.e. to the origin. Therefore its negative intersection with the plane $y=0$ occurs at positive values of $x$.

\begin{figure*}[!t]
\centering
\resizebox{\hsize}{!}{\includegraphics{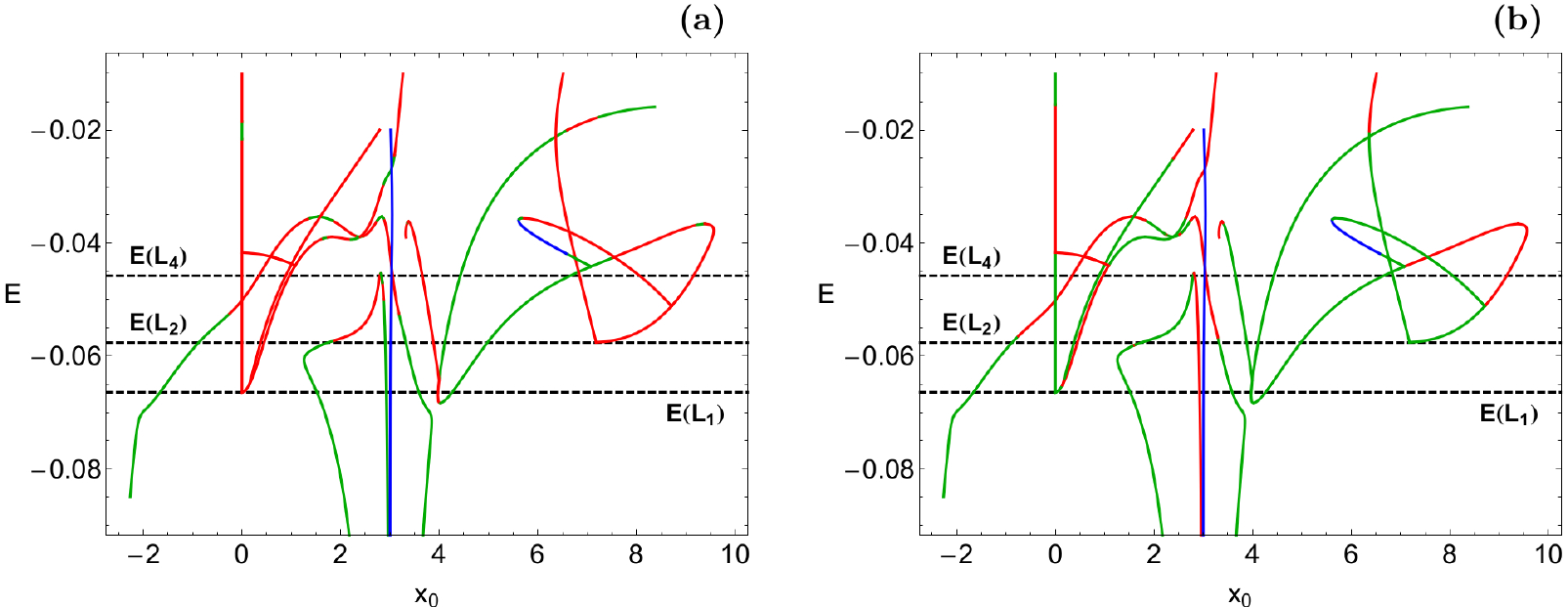}}
\caption{Evolution of (a-left): the normal and (b-right): the tangential stability of the periodic orbits, as a function of the total orbital energy $E$. Green color indicates stable motion, red color indicates unstable motion, while blue color corresponds to complex instability, where the monodromy matrix cannot be decomposed into two real $ 2 \times 2$ blocks. The horizontal black dashed lines correspond to the critical levels of the energy. (For the interpretation of references to colour in this figure caption and the corresponding text, the reader is referred to the electronic version of the article.)}
\label{trs}
\end{figure*}

The periodic orbits in position space near the potential well for $E = -0.06$ are plotted in Fig. \ref{orbsp2}. Note the following; All orbits with exception of mv2 are horizontal orbits, thus lying completely on the plane $z=0$. All orbits with the exception of mh2 (brown) and bb2 (orange) are approximately $x$-symmetric, with respect to the centre of the potential well, i.e. to the point $P_2$. However, mh2 moves already towards $L_1$ and bb2 moves towards $L_2$. This shows in which sense mh2 and bb2 look as if they would be a pair of orbits split off from ba2 in a pitchfork bifurcation. All orbits with the exception of br2a/br2b (pink) are exactly $y$-symmetric, whereas each individual orbit of the pair br2a/br2b breaks this symmetry, but the pair together is symmetric.

At the saddle energy $E(L_1)$ new important periodic orbits are born over the saddle point $L_1$. The saddle point $L_1$ is an index-1 saddle, therefore it has 2 stable modes and the two corresponding Lyapunov orbits. There is an orbit which oscillates in the $z$-direction exactly at $x=0$ and $y=0$, because of symmetry reasons, it is called lv1 and marked dark green. And there is an orbit circulating in the horizontal plane $z=0$ with negative orientation, it is called lh1 and marked red. Soon after its creation, near $E = -0.065$, lh1 suffers a pitchfork bifurcation where it splits off the two tilted loop orbits lt1a/lt1b, marked cyan. Near $E \approx -0.044$ lh1 suffers another pitchfork bifurcation where it splits off the pair of tilted loop orbits lu1a/lu1b, marked purple in the plot. These tilted loop orbits are split off from a horizontal orbit, however with increasing energy they become vertical extremely rapidly and near $E \approx -0.0415 $ they are absorbed by lv1 in a pitchfork bifurcation.

Near $E \approx -0.039$ very complicated events occur over the saddle $L_1$. To begin with, lh1 is destroyed in a saddle-centre bifurcation together with another horizontal orbit called bd2 near $E \approx -0.0389$. bd2 itself is created in a saddle-centre bifurcation near $E \approx -0.0394$ together with the orbit be2. Because bd2 and be2 are logical continuations of lh1, they are plotted in the same colour red as lh1. The total effect of these two saddle-centre bifurcations is to replace the orbit lh1 by the similar orbit be2 encircling the origin in a wider loop. Please be aware that the orbits lh1, bd2 and be2 are all symmetric relative to the origin, i.e. relative to the saddle $L1$. However, in the plot we mark the negative intersection with the plane $y=0$ and these intersections occur for positive values of $x$. Of course, the corresponding intersections with the plane $y=0$ with positive orientation occur at the corresponding negative values of $x$.

Near $E \approx -0.0389$ the orbit be2 suffers a pitchfork bifurcation where it split off the two orbits bf1 and bf2. Near
$E \approx -0.0353$ bf1 and bf2 destroy the orbits mh1 and mh2 respectively in saddle-centre bifurcations. The orbits bf1 and bf2 are the logical continuations of mh1 and mh2, therefore they are plotted in the same colour brown, as mh1 and mh2. The big global picture is that in the end the two horizontal orbits mh1 and mh2, with negative orientation coming out of the two potential holes, end in a pitchfork bifurcation where they are absorbed by the horizontal orbit with negative orientation coming from the saddle $L_1$.

The periodic orbits in position space near $L_1$ for $E = -0.043$ are plotted in Fig. \ref{orbsl1}. Panel (a) gives a three-dimensional perspective view, while panel (b) shows the projections into the 3 coordinate planes. Here the tilted loop orbits lu1a/lu1b (purple) are in the middle of their switch from horizontal orbits to vertical orbits. In addition we see how the pair mh1/mh2 has almost finished its transformation into orbits over the saddle $L_1$. Under a small further increase of energy they are squeezed flat in $x$-direction and approach the red orbit lh1 ready to disappear in a pitchfork bifurcation.

The saddle $L_2$ is also an index-1 saddle. Therefore at $E(L_2)$ two new Lyapunov orbits are born over $L_2$. One of them is a vertical orbit called lv2, it is plotted dark green. The other one is a horizontal orbit circling in negative orientation in the horizontal plane, let us call it lh2, it is plotted red. Near $E \approx -0.051$ lh2 suffers a pitchfork bifurcation where it splits off the pair lt2a/lt2b of tilted loop orbits, plotted cyan. Near$E \approx -0.044$ the orbit bb2 has already moved quite far away from the potential hole over $P_2$ so that it is near the saddle $L_2$ and at this energy also this orbits splits off a pair of tilted loop orbits, we will call them lu2a/lu2b, they are plotted purple. Near $E \approx -0.0366$ the orbits lh2 and bb2 collide and destroy each other in a saddle-centre bifurcation. Near $E \approx -0.0356$ also the pairs of tilted loop orbits lu2a/lu2b and lt2a/lt2b collide and destroy each other in two symmetry related saddle-centre bifurcations.

The periodic orbits in position space near $L_2$ for $E = -0.04$ are plotted in Fig. \ref{orbsl2}. Again panel (a) gives the three-dimensional perspective view, while panel (b) gives the projections into the 3 coordinate planes. Here bb2 (orange) has already finished its transformation into an orbit sitting over $L_2$. We also see how the tilted loop orbits from the pair lu2a/lu2b (purple) converge from the inside to the tilted loop orbits from the pair lt2a/lt2b (cyan). These two pairs are almost ready to collide and to destroy each other in a pair of symmetry related saddle-centre bifurcations.

In the next plot of Fig. \ref{trs} we give the stability properties of all these periodic orbits. First we calculated the 4-dimensional monodromy matrix of these orbits, then whenever it is possible we decompose them into two real 2-dimensional blocks. For such orbits belonging to a NHIM one of these blocks indicates the stability normal to the NHIM surface and the other one indicates the stability tangential to the NHIM surface. For the orbits which are not part of a NHIM we relate these two blocks to normal or tangential, according to their bifurcation properties with orbits belonging to a NHIM. Next we
calculate the traces of the two blocks of the monodromy matrix. These traces indicate the stability properties. If the trace is between -2 and +2 then we have stability, otherwise we have instability. The curves in the two panels of Fig. \ref{trs} are a repetition of the curves seen in Fig. \ref{fpo}, therefore a comparison with Fig. \ref{fpo} shows which curve is related to which periodic orbit. By colour we indicate the stability properties where red means unstable and green means stable. Panel (a) of Fig. \ref{trs} shows the parametric evolution of the normal stability properties, while panel (b) of the same figure shows the parametric evolution of the tangential stability.

The tilted loop orbits lu2a/lu2b are complex spiralling cases in the energy interval $E \in (-0.0419, -0.0358)$ and also mv2
is such a case. Then the monodromy matrix does not decompose into two real $ 2 \times 2$ blocks. Such cases are indicated by blue colour in Fig. \ref{trs}.

We do not describe the periodic orbits for energy values higher than -0.034 because for higher energies they are no longer
relevant for the escape over the saddles. And we will see that there are also no interesting events on the NHIMs for still
higher values of the total orbital energy.

\subsection{The restricted Poincar\'e map}
\label{ss3}

The NHIM surface $S_E$ in the map is an invariant surface, i.e. when we choose some initial point from this surface, then also all images and all pre-images of this point lie on $S_E$. Therefore the restriction of the Poincar\'e map $P_{res,E}$ to $S_E$ exists and its plots are the best presentation of the development scenario of the NHIM. These plots automatically include the development scenario of the Lyapunov orbits and of some other important periodic orbits related to the Lyapunov orbits and belonging to the NHIM.

The numerical plots of $P_{res}$ are done along the following lines. For an energy very close to the saddle energy we know $S_E$ from the analytical calculations in the quadratic approximations. Let us assume we have it for an energy value $E_1$. Next we want to construct $S_E$ and $P_{res,E}$ for the energy value $E_2$ a little larger. Then $S_{E_2}$ lies in the domain of the map close to the region occupied by $S_{E_1}$. So we cut out a neighbourhood $N_E$ around this region and choose a line $L_i$ of initial conditions in $N_E$. A line in general position will intersect the local segment of the stable manifold $W^s(S_{E_2})$ of $S_{E_2}$ in a single point. Now we run many orbits starting on $L_i$ and measure the time such orbits stay inside of $N_E$. The points close to the intersection with $W^s(S_{E_2})$ have high values of this time. Now we can zoom in on this subsegment with large times and come closer to the intersection point within our numerical accuracy. Then we store a few points of this orbit and repeat the search for large dwelling times with the next point, etc.
So we keep the numerical orbit close to an unstable invariant subset. In essence we combine the 4-dimensional Poincar\'e map with a projection onto the local segment of the stable manifold of the NHIM. This is a version of the control of chaos \cite{OGY90,SOGY90}. Thereby we obtain simultaneously first many points lying on $S_E$ and second we obtain the restricted map. More details on the construction of $P_{res}$ are given in \cite{GDJ14} while some further examples of this construction can be found in \cite{GJ15,JZ16,ZJ17}.

\begin{figure*}[!t]
\centering
\resizebox{\hsize}{!}{\includegraphics{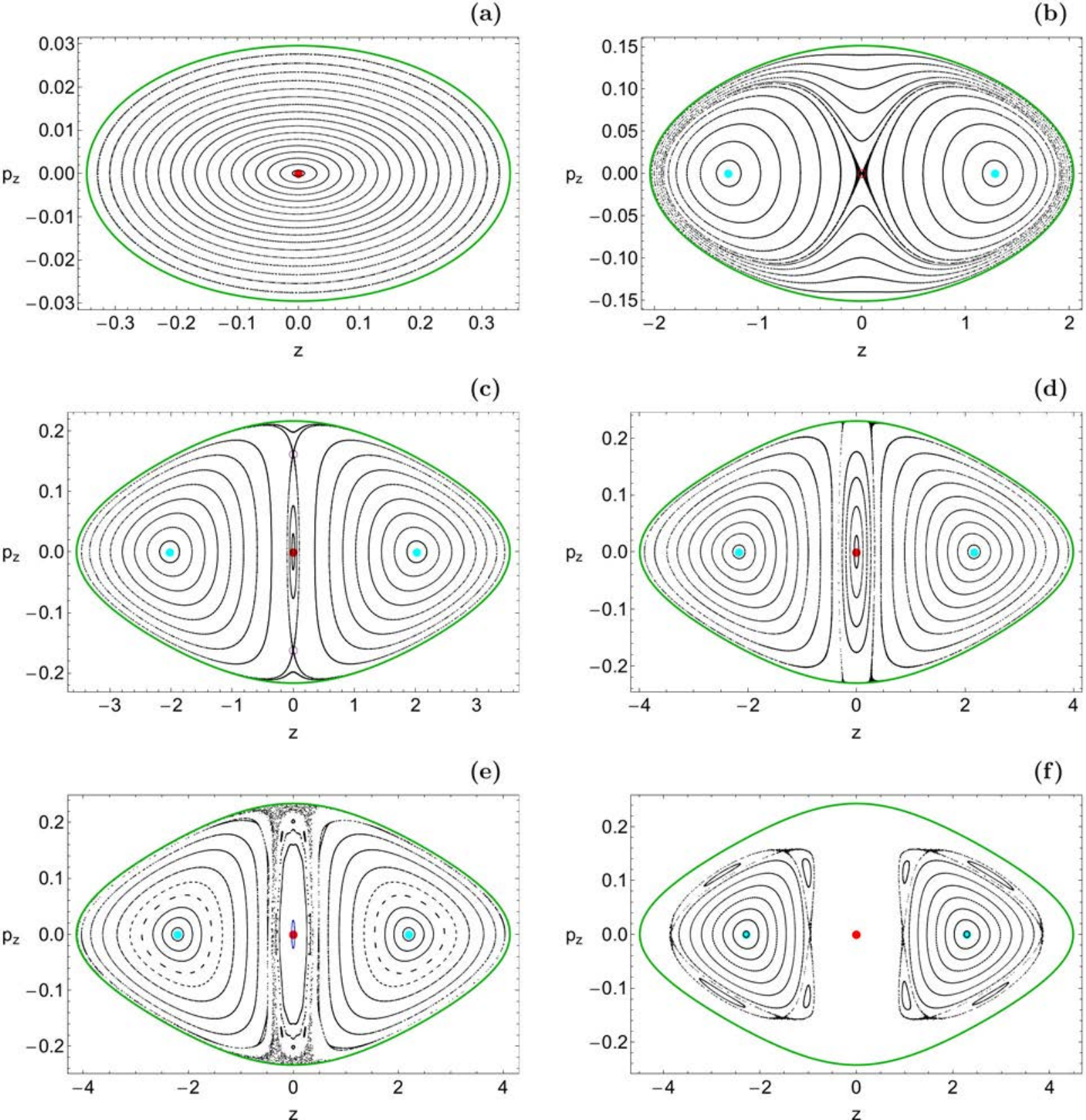}}
\caption{Projections of the NHIM surfaces, over $L_1$, into the $(z,p_z)$ plane. The outermost green solid closed curve corresponds to the vertical Lyapunov orbit. (a): $E = -0.066$; (b): $E = -0.055$; (c): $E = -0.043$; (d): $E = -0.040$; (e): $E = -0.03915$; (f): $E = -0.037$. The colour code is the same as in Fig. \ref{fpo}. (For the interpretation of references to colour in this figure caption and the corresponding text, the reader is referred to the electronic version of the article.)}
\label{mapsl12}
\end{figure*}

\begin{figure*}[!t]
\centering
\resizebox{0.8\hsize}{!}{\includegraphics{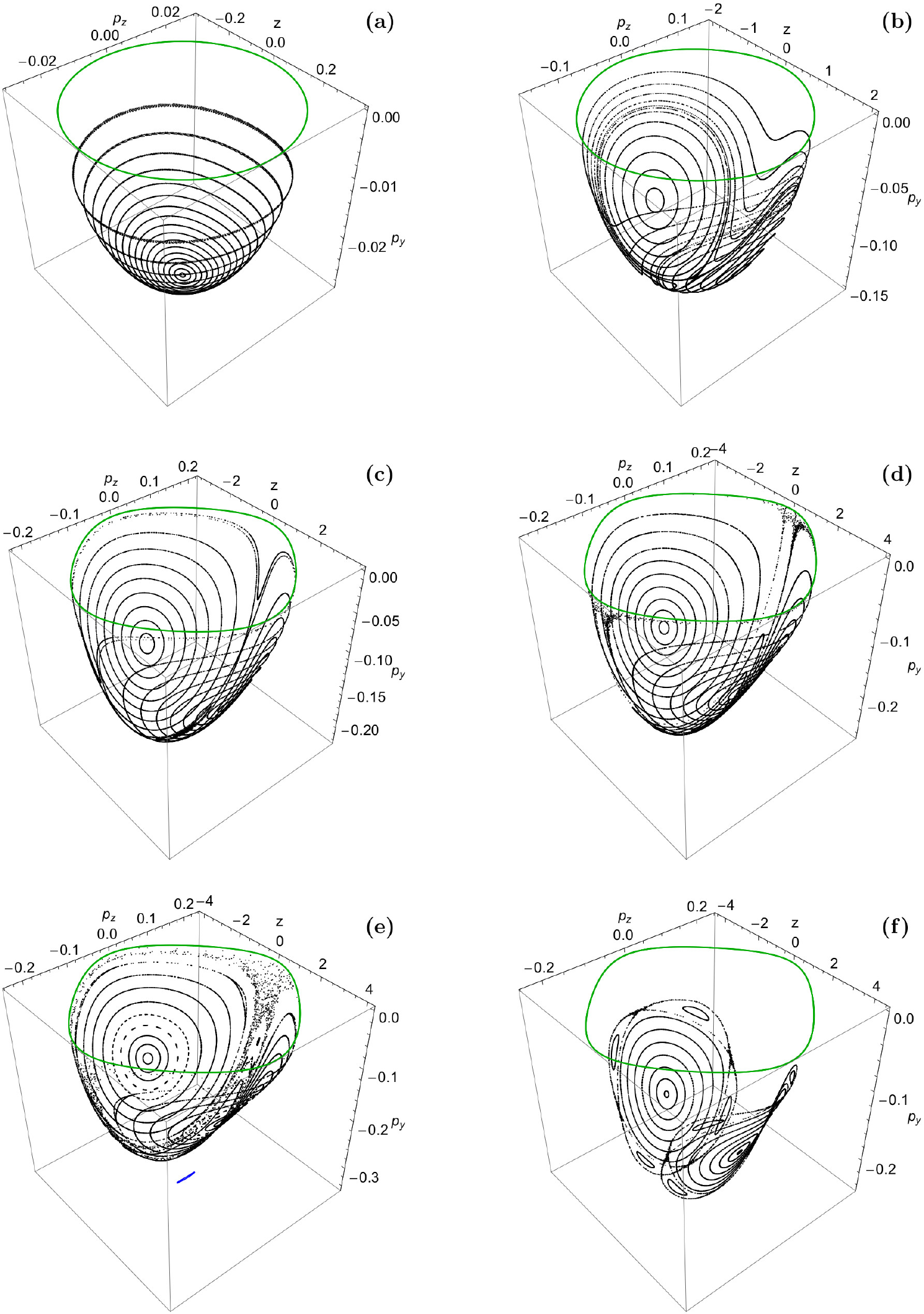}}
\caption{Projections of the NHIM surfaces, over $L_1$, into the $(z,p_y,p_z)$ phase space. (a): $E = -0.066$; (b): $E = -0.055$; (c): $E = -0.043$; (d): $E = -0.040$; (e): $E = -0.03915$; (f): $E = -0.037$. The colour code is the same as in Fig. \ref{fpo}. (For the interpretation of references to colour in this figure caption and the corresponding text, the reader is referred to the electronic version of the article.)}
\label{mapsl13}
\end{figure*}

\begin{figure*}[!t]
\centering
\resizebox{\hsize}{!}{\includegraphics{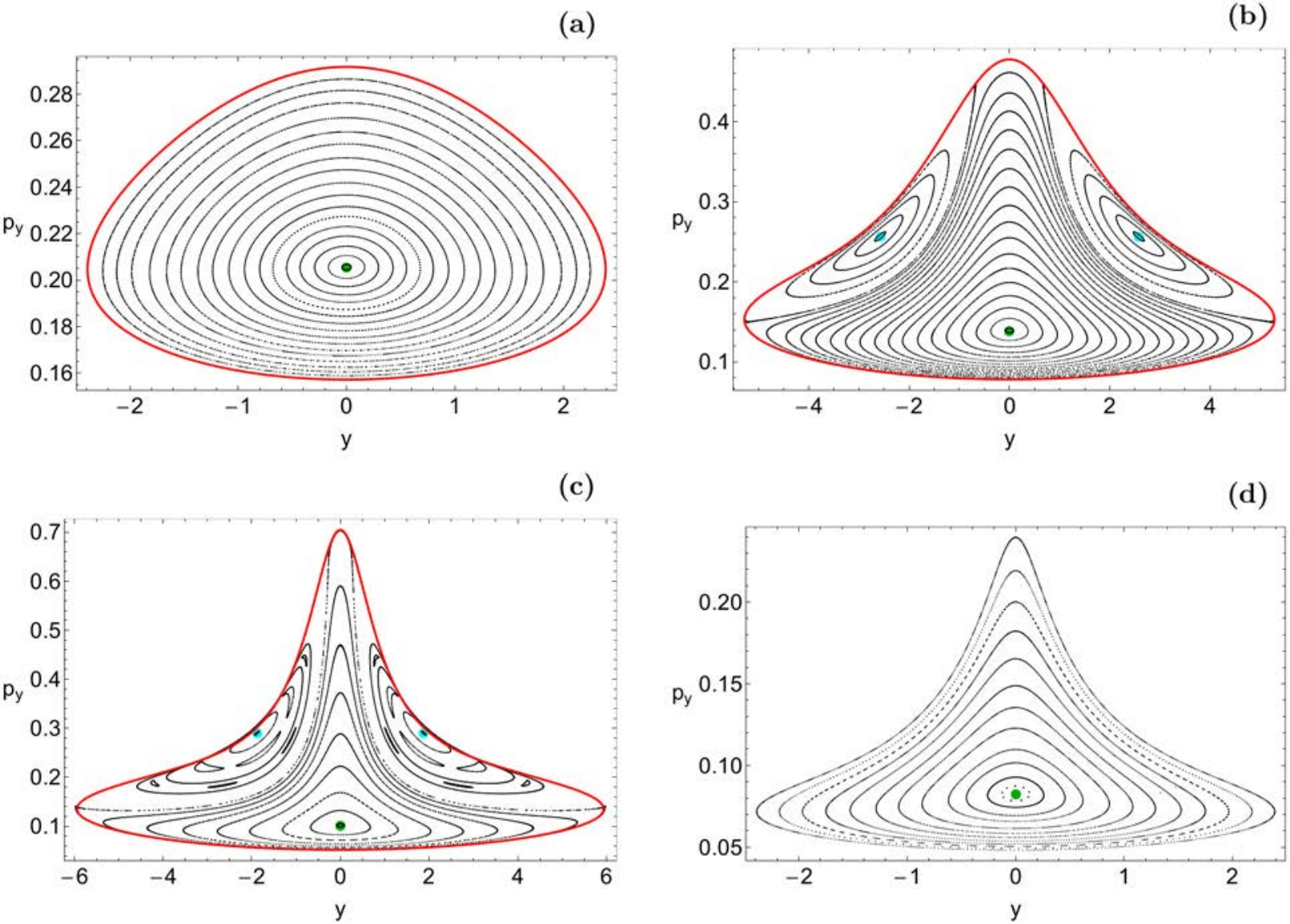}}
\caption{Projections of the NHIM surfaces, over $L_2$, into the $(y,p_y)$ plane. The outermost red solid closed curve corresponds to the horizontal Lyapunov orbit. (a): $E = -0.055$; (b): $E = -0.043$; (c): $E = -0.037$; (d): $E = -0.034$. The colour code is the same as in Fig. \ref{fpo}. (For the interpretation of references to colour in this figure caption and the corresponding text, the reader is referred to the electronic version of the article.)}
\label{mapsl22}
\end{figure*}

\begin{figure*}[!t]
\centering
\resizebox{\hsize}{!}{\includegraphics{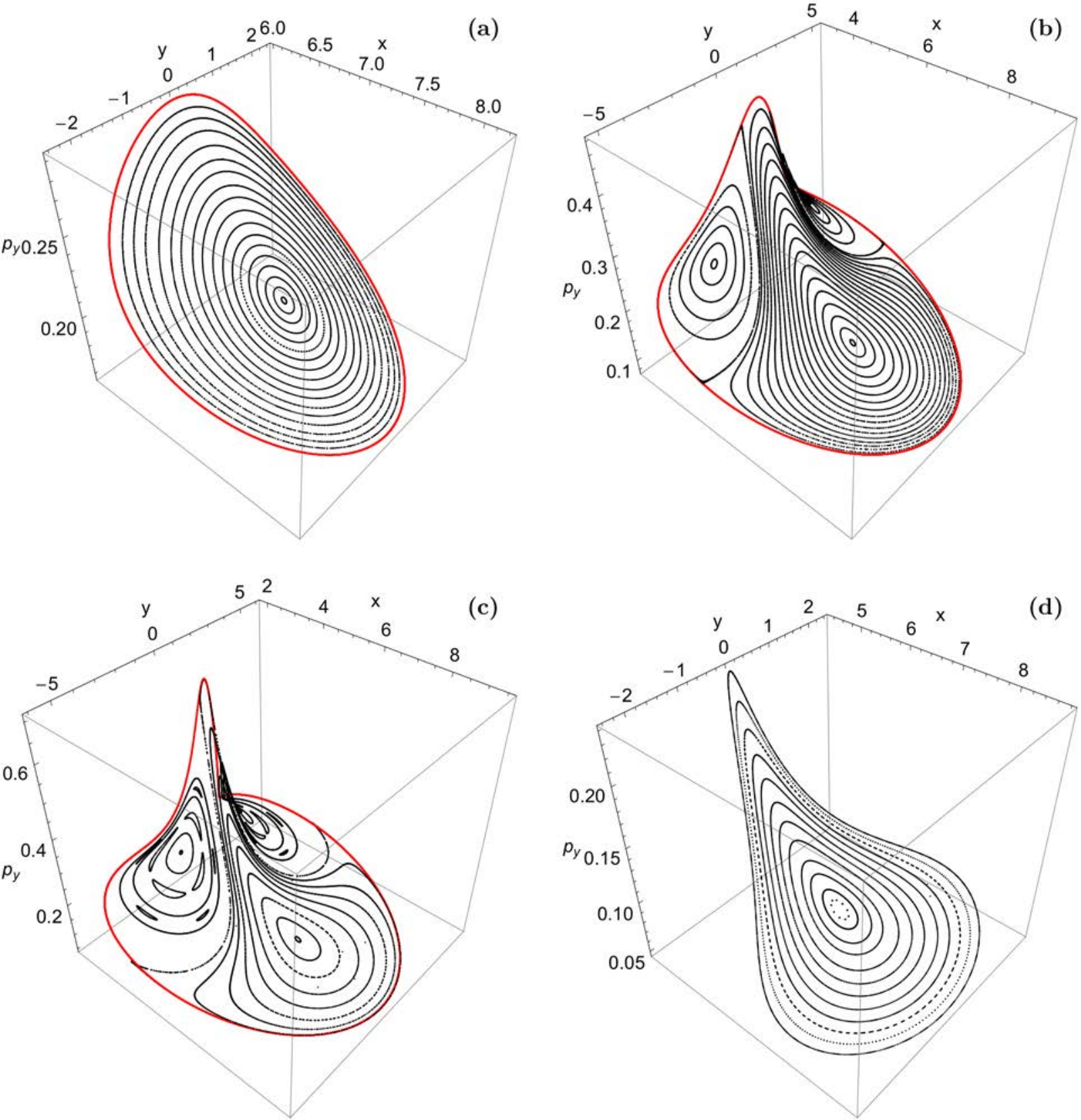}}
\caption{Projections of the NHIM surfaces, over $L_2$, into the $(x,y,p_y)$ phase space. (a): $E = -0.055$; (b): $E = -0.043$; (c): $E = -0.037$; (d): $E = -0.034$. The colour code is the same as in Fig. \ref{fpo}. (For the interpretation of references to colour in this figure caption and the corresponding text, the reader is referred to the electronic version of the article.)}
\label{mapsl23}
\end{figure*}

These invariant surfaces are 2-dimensional surfaces in a 4-dimensional embedding space, namely in the domain of the full 4-dimensional Poincar\'e map. To give some graphical impression of this embedding we show 2 different views of each NHIM surface. First we project it into either the $(y,p_y)$ coordinate plane or into the $(z,p_z)$ coordinate plane and in addition we show a perspective view of a projection into either the $(x,y,p_y)$ space or into the $(z,p_y,p_z)$ space, respectively. The combination of the 2 different representations gives at least a slight impression of the complicated curvature of the surfaces in the 4-dimensional space which develops when the energy increases.

Next let us follow the development scenario over the saddles in the NHIM plots. First we begin with the NHIM over $L_1$ and let us call this one NHIM1. The most convenient intersection condition for the restricted map on NHIM1 is $y=0$ with negative orientation. The Lyapunov orbit lv1 lies completely within this surface, therefore this orbit will act as the boundary of the domain of $P_{res}$. In the domain we use the canonical coordinates $z$ and $p_z$. Panel (a) of Fig. \ref{mapsl12} shows the restricted map for $E = -0.066$. Here we are still close to the quadratic approximation and the plot looks like the Poincar\'e map of a 2-dof anisotropic harmonic oscillator. As already mentioned the boundary is lv1, in the plot it is the dark green line. If we would prefer to represent this fundamental orbit by a fixed point of the map, then we could pinch the boundary together to a point and thereby change the domain of the map from a disc to a sphere. However it is easier to plot a disc. The orbit lh1 intersects the plane perpendicular and the corresponding fixed point of the map is the center of the plot, and in the representation in the $(z,p_z)$ plane it is marked by a filled red circle, the same colour as the periodic
orbit lh1 in the bifurcation diagram and in the orbit plot, shown in Figs. \ref{fpo} and \ref{orbsl1}, respectively. Because of symmetry reasons this fixed point always lies in the origin of the coordinate system. The invariant lines in the plot represent quasi-periodic superpositions of the two stable modes of motion over the saddle $L_1$ where the energy in the horizontal mode changes from its maximally possible value in the middle to 0 on the boundary. Both fundamental periodic orbits are stable in the tangential direction to the NHIM1. However, it is seen that all structures are highly unstable in the directions normal to the NHIM surface.

In panel (b) of Fig. \ref{mapsl12} we increase the energy to the value -0.055. Here the orbit lh1 has already suffered its pitchfork bifurcation and has become unstable in tangential direction. However, the tangential instability is small compared to the normal instability. Therefore the orbit is still normally hyperbolic and remains a part of the NHIM. Because the fixed point is unstable it is now marked by an open red circle. Now the central point in the map is the center of a small chaos strip which still looks like a separatrix in the numerical plot. The two tilted loop orbits are tangentially stable and are represented by the two new fixed points which are the centres of the islands enclosed by the separatrix. In the plot in the $(z,p_z)$ plane they are marked by filled cyan circles, the same colour as the corresponding orbits in the orbit plots.

In panel (c) of the same figure we have proceeded to the energy -0.043. Now the central point of the map, i.e. the orbit lh1, has suffered its second pitchfork bifurcation where it splits off the pair lu1a/lu1b of tilted loop orbits. The central fixed point has split off two fixed points which are unstable in tangential direction and the centre has returned to tangential stability. As a stable fixed point it is again marked by a red filled circle. The two split off fixed points representing the tilted loop orbits lu1a/lu1b have become the central points of the separatrix (fine chaos strip). As unstable fixed points they are marked by open circles and in purple colour, the same colour as the corresponding orbits in the orbit plots.

The next panel (d) of Fig. \ref{mapsl12} show the situation for $E = -0.04$. Now the boundary, i.e. lv1, has run through the pitchfork bifurcation where it absorbed the pair of tilted loop orbits lu1a/lu1b. Thereby the boundary curve has become unstable in tangential direction. Here its tangential instability is still small compared to its normal instability. So it remains a part of the NHIM, the boundary in our representation.

In panel (e) of the same figure we have the case of $E = -0.03915$, i.e. we are in the small energy interval where the orbit bd2 exists. Now the orbit lh1 has become normally elliptic and can no longer be a part of the NHIM1. Therefore a hole has been created in the middle of the surface. This hole is still surrounded by a KAM line which separates the hole from the chaos strip grown out of the separatrix. Therefore the orbits from the chaos strip are still restricted from entering the hole and being lost by falling over the inner edge of the NHIM. On the other hand, at this energy the orbit bd2 exists and it is normally hyperbolic. Therefore this corresponding fixed point is now marked by the filled red circle in the middle. There has been created a very small piece of invariant surface around this new fixed point. One KAM curve coming from this new piece of invariant surface is included in the plots. To distinguish it well from the old NHIM surface, the new piece is shown in blue colour. In the projection into the $(z,p_z)$ plane this new fixed point and its neighbourhood replace exactly lh1 and its neighbourhood. However, in the perspective view we see that the new piece of surface lies well outside of the old surface.

Finally in panel (f) of Fig. \ref{mapsl12} the energy has increased to -0.037. Now the chaos strip grown out of the separatrix has disappeared from the invariant surface. Also the orbit lv1 belongs to this chaos strip and does no longer belong to the NHIM surface for this energy. Accordingly the only surviving pieces of the invariant surfaces are the neighbourhoods of the tilted loop orbits lt1a/lt1b. These orbits and their neighbourhoods remain normally hyperbolic and remain invariant surfaces up to rather high energies. However, for high energies such invariant surfaces are no longer important for the flow over the saddle. Therefore we do present the further scenario of these surfaces for still higher energies. The central filled red circle in the representation in the $(z,p_z)$ plane corresponds now to the orbit be2 which has replaced lh1. However, as should be evident from the plot, this fixed point is now longer part of the NHIM surface.

The corresponding plots in Fig. \ref{mapsl13} of the perspective views in the $(z,p_y,p_z)$ space give an impression how the invariant surface in the higher dimensional domain of the full map has a bowl shape. In order not to overload these perspective views we did not mark the fixed points, as in Fig. \ref{mapsl12}.

Now we come to the NHIM over the saddle $L_2$, let us call it NHIM2. Here it was most convenient to use the intersection condition $z = 0$ for the Poincar\'e map. In Figs. \ref{mapsl22} and \ref{mapsl23} we plot the restricted map on this surface in the $(y,p_y)$ plane and in a perspective view in the $(x,y,p_y)$ space, respectively. In panel (a), where $E = -0.055$, $P_{res}$ comes very close to a Poincar\'e map for a 2-dof anisotropic harmonic oscillator. The central fixed point represents the vertical Lyapunov orbit lv2 and in the representation in the $(y,p_y)$ plane it is represented by a solid dark green circle, the colour of the corresponding orbit in the orbit plots. The boundary of the NHIM2 is the horizontal Lyapunov orbit lh2, it is plotted red equal as in the orbit plots. Again the KAM curves represent quasi-periodic superpositions of the motion along the two stable modes of motion over $L_2$.

In panel (b) of Figs. \ref{mapsl22} and \ref{mapsl23} the energy has increased to -0.043. Now the boundary (the orbit lv2) has already suffered its pitchfork bifurcation where it has split off the tilted loop orbits lt2a/lt2b and has become unstable. The tilted loop orbits show up as new fixed points of $P_{res}$. In the representation in the $(y,p_y)$ plane they are marked by solid cyan circles, the same colour as the corresponding orbits in the orbit plots.

Panel (c) of the same figures shows the situation at energy -0.037. Compared to panel (b) no qualitative changes on large scale have occurred. The perspective view shows how curved the surface has become in the higher dimensional embedding space and this makes it understandable that the whole construction has already come close to structural instability. If we increase the energy only a little bit more, than first the boundary together with the separatrix (small chaos strip) looses the normal hyperbolicity and is lost from the invariant surface. For a small further increase of the energy the tilted loop orbits are destroyed in their saddle-centre bifurcation and then also the islands around the tilted loop orbits are lost.

All what remains for higher energies is shown in panel (d) of the same figures for $E = -0.034$. It is a small neighbourhood around the vertical orbit lv2. This piece of invariant surface survives up to high energies but is no longer important for the escape over the saddle $L_2$. Therefore we do not follow its further development.

Of course, there is another NHIM over the saddle $L_3$, let us call it NHIM3. However, this one is nothing really new, because we can obtain it from NHIM2 by a rotation by $\pi$ around the vertical $z$-axis.

\subsection{The stable and unstable manifolds of the NHIMs}
\label{ss4}

\begin{figure*}[!t]
\centering
\resizebox{\hsize}{!}{\includegraphics{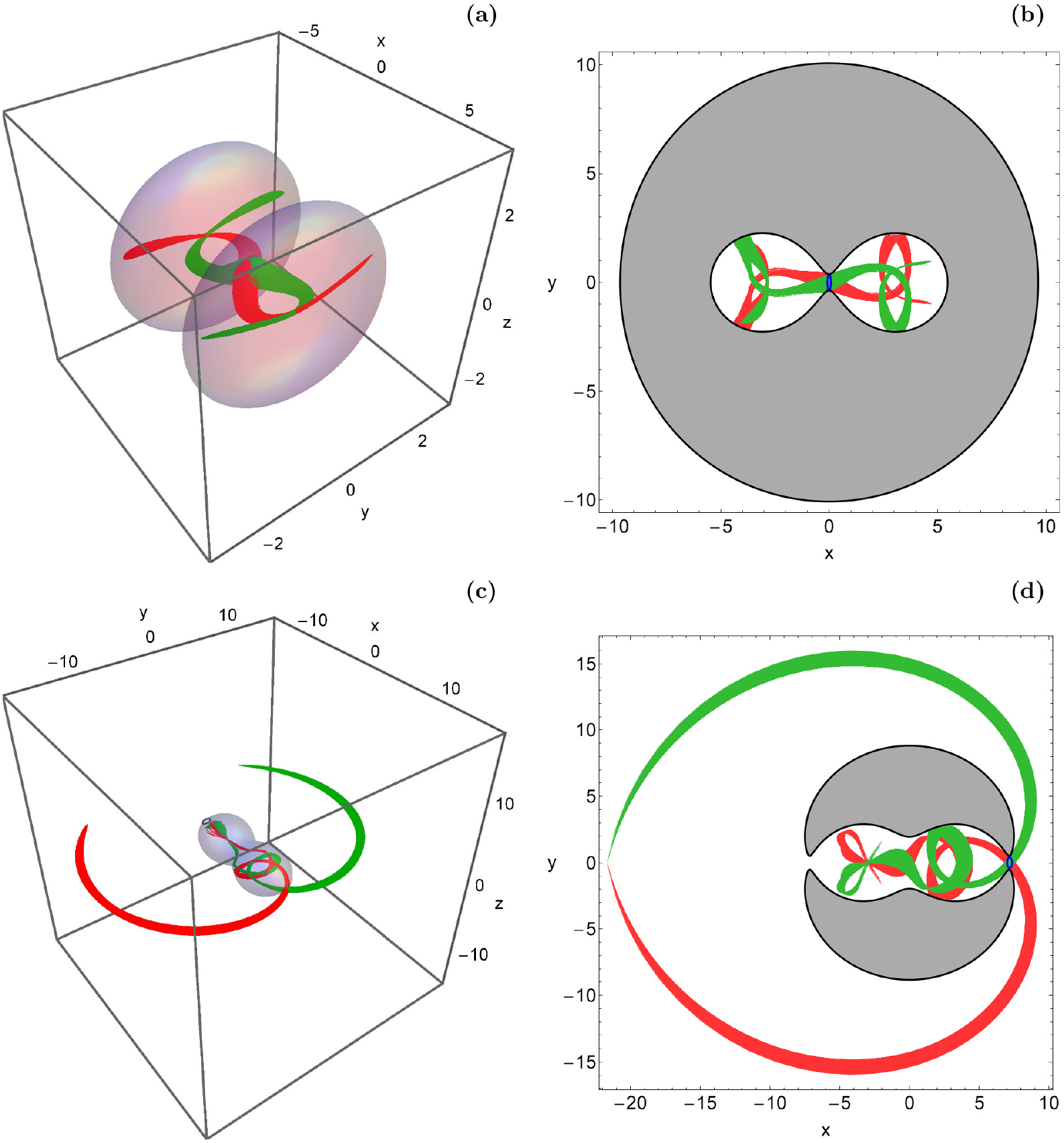}}
\caption{The perspective plots of the stable manifold $W^s(S_E)$ (green) and the unstable manifold $W^u(S_E)$ (red), when (a-upper left): $E = -0.066$ and (c-lower left): $E = -0.0575$. Panels (b) and (d): The corresponding projections of the stable and the unstable manifolds on the configuration $(x,y)$ space. The corresponding horizontal Lyapunov orbits are shown in blue. (For the interpretation of references to colour in this figure caption and the corresponding text, the reader is referred to the electronic version of the article.)}
\label{mans}
\end{figure*}

The importance of the NHIMs over the saddles and of their stable and unstable manifolds is based on the following behaviour of general orbits. If the energy of the particle is a lot larger than the saddle energy, then the orbit crosses the saddle without problems and leaves the potential interior rather fast. Therefore such particles have left the potential region already long time ago. Really interesting are particles which start with an energy a little below the saddle energy. As long as the energy of such particles is conserved, they are not able to cross the saddle. However, from time to time such particles suffer close encounters with other objects in the potential interior and then there might be energy exchange between the particles and the particle under observation might increase its energy a little and result in an energy just a little above the saddle energy. Then for exactly such particles the NHIM over the saddle and its stable and unstable manifolds become really important. We denote the stable and unstable manifolds of the NHIM by $W^s(NHIM)$ and $W^u(NHIM)$, respectively. Both have two branches, one going to one side of the saddle and the other one going to the other side of the saddle.

After some while the particle with an energy a little above the saddle energy can come close to the local segment of
$W^s(NHIM)$ and then it approaches the saddle along this stable manifold and spends some time over the saddle region. The further fate of the particle depends on which side of $W^s(NHIM)$ the particle has come in. If it has come in on one side of $W^s(NHIM)$, then it will later leave the saddle region along one of the two branches of $W^u(NHIM)$, if it has come in on the other side of $W^s(NHIM)$ then it leaves the saddle region along the other branch of $W^u(NHIM)$. In this sense the local branch of $W^s(NHIM)$ is the dividing surface in phase space between going over the saddle and returning to the same side of the saddle. In the end the NHIM and its manifolds guide the transport over the saddle and the positions of these manifolds indicate from which direction particles approach the saddle regions and into which directions they leave it again.

These explanations give the motivation for the next figure \ref{mans}. It shows the projection into the position space of the local segments of the stable and unstable manifolds of NHIM1 and of NHIM2. Stable manifolds are plotted in green, while unstable manifolds are plotted red. Panels (a) and (b) show the manifolds of NHIM1 for $E = -0.066$, while parts (c) and (d) of the same figure show the manifolds of NHIM2 for $E = -0.0575$, i.e. always for energies just a little above the respective saddle energy. Panels (a) and (c) show a perspective view in the 3-dimensional $(x,y,z)$ position space, while panels (b) and (d) show a projection into the configuration $(x,y)$ plane. To produce the plots we select a lot of initial conditions very close to the NHIMs, we obtain them essentially from the plots in Figs. \ref{mapsl12} and \ref{mapsl22}. Then we let the orbits for these initial conditions run for a finite time\footnote{The total time of the numerical integration in each case was chosen such that the final shape of the manifolds is as clear as possible.} as well into the future as into the past. Such orbits converge automatically towards $W^u(NHIM)$ in the future and towards $W^s(NHIM)$ in the past. These orbits have a random distribution over the two branches of the manifolds. The horizontal Lyapunov orbits over the saddles are included as blue loop. The black curves in panels (b) and (d) are the boundaries of the energetically accessible parts of the position space in the $(x,y)$ plane.

In panel (b) of Fig. \ref{mans} we see the flow pattern between the two potential holes for an energy just above $E(L_1)$. Moreover in panel (d) we see the flow over the saddle $L_2$. Here note that the energy is already way above $E(L_1)$ and therefore the inner branches of $W^{s/u}(NHIM2)$ cross the saddle $L_1$ without any problems. Of course, the manifolds of NHIM3 are obtained from the ones of NHIM2 by a rotation around the vertical $z$-axis by $\pi$.

\subsection{A tentative classification of the NHIM scenarios found so far}
\label{ss5}

A general theory for bifurcations and development scenarios of NHIMs is an open problem. Therefore we are not yet able to place our example into some general theoretical framework. The interested reader can find some information on this fascinating topic in \cite{AB12,LST06,MS14,MCE13,TTK11,TTK15,TTT15}. One possibility to advance in this problem might be to study many examples of development scenarios and to look for repeating patterns. We consider our present investigation a contribution to this programme.

In the present subsection we summarise the most important properties of the scenarios found in the present system, the binary dwarf galaxy system (abbreviated bdg-system in the following) and compare them to similar scenarios found in other systems also belonging to the same wide class of systems from celestial mechanics being treated in the rotating frame. The NHIM scenarios are best explained by the properties and bifurcations of the Lyapunov orbits, because these Lyapunov orbits are the skeletons of the NHIMs and determine their scenario to a large extent.

\begin{itemize}
  \item \textbf{Case A}:

  First, we have found a scenario which we call case A and which has shown up in a model for barred galaxies \cite{JZ16}. For energies going up from the saddle energy the scenario starts with a pitchfork bifurcation of the vertical Lyapunov orbit where it splits off a pair of tilted loop orbits and becomes tangentially unstable. This event if followed very soon by a second pitchfork bifurcation of the vertical Lyapunov orbit where it splits off another pair of tilted loop orbits and returns to tangential stability. Here the horizontal Lyapunov orbit does not take part in the bifurcation scenario up to very high energies where it finally looses its normal hyperbolicity and is lost from the NHIM. The scenarios of the horizontal and the vertical Lyapunov orbits are never linked together in this case.

  Recently we have made some so far unpublished preliminary calculations for a model potential leading to four index-1 saddles of the effective potential where one pair of saddles is nonequivalent to the other pair. However, all the saddle neighbourhoods are similar to the ones in the barred galaxy. Also here the scenario A is realised in essentially the same form over all the four index-1 saddles.

  This case is favoured by the following properties of the saddle dynamics. Exactly at the saddle energy the horizontal period is larger than the vertical period. However with increasing energy the horizontal period first becomes smaller and only for rather large energy it increases again. So in the horizontal motion a positive anharmonicity dominates, the frequency first increases. This mode is stiff. In contrast the vertical motion has a strong negative anharmonicity, with increasing energy the period increases, i.e. the frequency decreases. This mode is very floppy. Accordingly for an energy relatively little above the saddle energy the two modes come into a 1:1 resonance when the vertical period passes the horizontal period. At the same time the tangential trace of the monodromy matrix of the vertical Lyapunov orbit changes rapidly with increasing energy and a little before the 1:1 resonance it crosses the value 2 from below and then it turns around and a little above the 1:1 resonance it passes the value 2 again, this time from above. This explains the 2 pitchfork bifurcations and the creation of two pairs of tilted loop orbits. In contrast the tangential trace of the horizontal orbit stays away from the value 2 for all energies near the saddle energy.

  \item \textbf{Case B}:

  Second, we have found a scenario which for the moment we call case B and which has shown up in the NHIM1 of the bdg-system. The scenario starts with a pitchfork bifurcation of the horizontal Lyapunov orbit. The next important bifurcation is another pitchfork bifurcation of the horizontal Lyapunov orbit where it splits off another pair of tilted loop orbits. These new tilted loop orbits turn into the vertical direction very rapidly and are absorbed by the vertical Lyapunov orbit in an inverse pitchfork bifurcation (see Fig. \ref{fpo}). Thereby, in case B the bifurcation scenarios of the horizontal and the vertical Lyapunov orbits are linked together, an event which we did not find in any other example so far. For still higher energy the horizontal Lyapunov orbit absorbs in an inverse pitchfork bifurcation two symmetrically placed orbits coming from the two potential minima. This event is only possible because of the left/right symmetry of the system and the symmetry enforced position of the NHIM1 in the middle, i.e. on a symmetry line. Therefore this part of the scenario is probably non-generic in the absence of a corresponding discrete symmetry.

  We are aware that probably also a part of the other properties of the bdg-system are caused by this discrete symmetry, which enforces the NHIM1 to lie in the plane $x = 0$. Therefore, as next step in the programme it would be natural to study a system with two (or more) nonequivalent NHIMs, where none of the NHIMs is enforced to have such a planar structure or a symmetrical position.

  Some properties of the saddle dynamics which makes case B possible are the following: At the saddle energy the horizontal period starts with a value a little smaller than the vertical period. However, the horizontal period increases rapidly so that it passes the vertical period for an energy close to -0.05. In contrast, the vertical period increases rather slowly with increasing energy. The resulting 1:1 resonance between the horizontal and the vertical frequency again leads to a pair of pitchfork bifurcations, this time performed by the horizontal Lyapunov orbit.

  The common pattern among the cases A and B is as follows: The period which start as the smaller one at the saddle energy is the one which increases faster with increasing energy and passes the other period soon. This 1:1 resonance induces a pair of pitchfork bifurcations in that Lyapunov orbit which belongs to the period with the faster speed of change. This Lyapunov orbit is the more structurally unstable one and the one which has to suffer the consequences of the resonance.

  \item \textbf{Case C}:

  Third, we have several examples of a further case which we call case C at the moment. Here the scenario starts with a pitchfork bifurcation of the horizontal Lyapunov orbit, where as usual a pair of tilted loop orbits is split off. However, this time the initial pitchfork bifurcation is not followed by a second one. This case is realised by NHIM2 in the bdg-system and it has also been observed by Jorba and Masdemont \cite{JM99}, for the standard three-body problem and by us \cite{ZJ17} in a model for the decay of a star cluster. In \cite{JM99} they studied the dynamics restricted to the center manifold of the saddle. But this is essentially the same idea as the restriction of the dynamics to the NHIM surface. Because the treatment used in \cite{JM99} is perturbative, unfortunately it does not show the events for high perturbations (high energy) and we can not include this system into the considerations for higher energies already far from the saddle energy.

  In the star cluster and in the NHIM2 of the bdg-system the vertical Lyapunov orbit never participates in any bifurcation scenarios. And in both of these systems at some high energy the horizontal Lyapunov orbit is destroyed in a saddle-centre bifurcation when it collides with a horizontal orbit growing out of a potential minimum. Also the tilted loop orbits which split off from the horizontal Lyapunov orbit are destroyed in related saddle-centre bifurcations when they collide with other tilted loop orbits split off from the horizontal orbits growing out of the potential minima. This scenario is in total rather simple and it gives the impression that the vertical degree of freedom is almost decoupled from the horizontal motion.

  In the star cluster model the periods of both Lyapunov orbits have an extremely small dependence on the energy only. Also the dependence of the tangential traces on the energy is rather weak, and these traces run into the value 2 for very high energy only. Accordingly, the pitchfork bifurcation of the horizontal Lyapunov orbit happens for an energy value which is already rather irrelevant for the saddle dynamics. The closeness of the whole system to harmonic behaviour may have the following explanation: The tidal part of the potential which is the important contribution near the saddle is given in quadratic approximation (see Eq. (2) in \cite{ZJ17}). The only contributions of higher orders come from the nucleus potential. But the nucleus potential is already rather flat in the saddle region, so that it is not able to produce significant anharmonic effects. This in turn avoids important resonances and avoids any complicated scenario. A similar argument may apply to the perturbative treatment of the three-body system in \cite{JM99}, where the only important bifurcation event in the whole scenario is the initial pitchfork bifurcation of the horizontal Lyapunov orbit (see figures 3 and 6 in \cite{JM99}).
\end{itemize}

\subsection{Some further comments on the development scenario}
\label{ss6}

Of course, we expect the existence of many more possible scenarios besides the ones mentioned in the previous subsection and also combinations of these scenarios might be common.

In all the scenarios there is always the possibility to have some of the fundamental orbits replaced by a similar one at a
shifted position as it has been described for the orbits lh2/bd2/be2 in subsection \ref{ss2} of the present publication for the bdg-system, belonging to case B. Such a replacement always runs through a pair of saddle-centre bifurcations. In the barred galaxy system, belonging to case A, this event occurs for the vertical Lyapunov orbit when we have sufficiently large values of the angular velocity $\Omega_{\rm b}$. The inclusion of such an event does not change the basic scenario.

In the bdg-system we have noticed a property which has called our attention: The invariant surfaces NHIM1 and NHIM2 are not related by any discrete symmetry and therefore at first sight there is no logical need that their development scenarios are strongly related. The same holds when we replace NHIM2 by NHIM3. Nevertheless, we observe in the bifurcation diagram of Fig. \ref{fpo} a coincidence which arises interesting questions. Around $E = -0.036$ there is an accumulation of bifurcations of elementary periodic orbits which needs some explanation. Over the saddle $L_2$ the orbits lh2 and bb2 disappear in a saddle-centre bifurcation. And also the pairs of tilted loop orbits lu2a/lu2b and lt2a/lt2a disappear in saddle-centre bifurcations. It is not so surprising that tilted loop orbits are destroyed at similar values of $E$ as their parent orbits. This has also happened in the development scenario of the star cluster. However, it is a lot more surprising that at nearby energies also important bifurcation events happen over the central saddle $L_1$. And even the pair br2a/br2b disappears around the same energy value even though this orbit pair has nothing to do with the index-1 saddles of the effective potential and with the NHIMs sitting over them. And as a consequence of these coincidences also all the NHIM surfaces suffer the most important decay processes in this relatively small energy interval.

An interesting speculation might be the following: There are heteroclinic connections between the various NHIMs in the system. And also further periodic orbits might be included into this global tangle. Along these heteroclinic connections
the NHIMs create mutual influences of their development scenarios and somehow synchronize their development scenarios and transmit tendencies for the creation of chaos and of decay. A deeper investigation of the heteroclinic structures in this system will therefore be a highly interesting and worthwhile project for the future and may clarify at least some of these mysteries.

\section{Stellar formations of escaping stars}
\label{spr}

\begin{figure}[!t]
\includegraphics[width=\hsize]{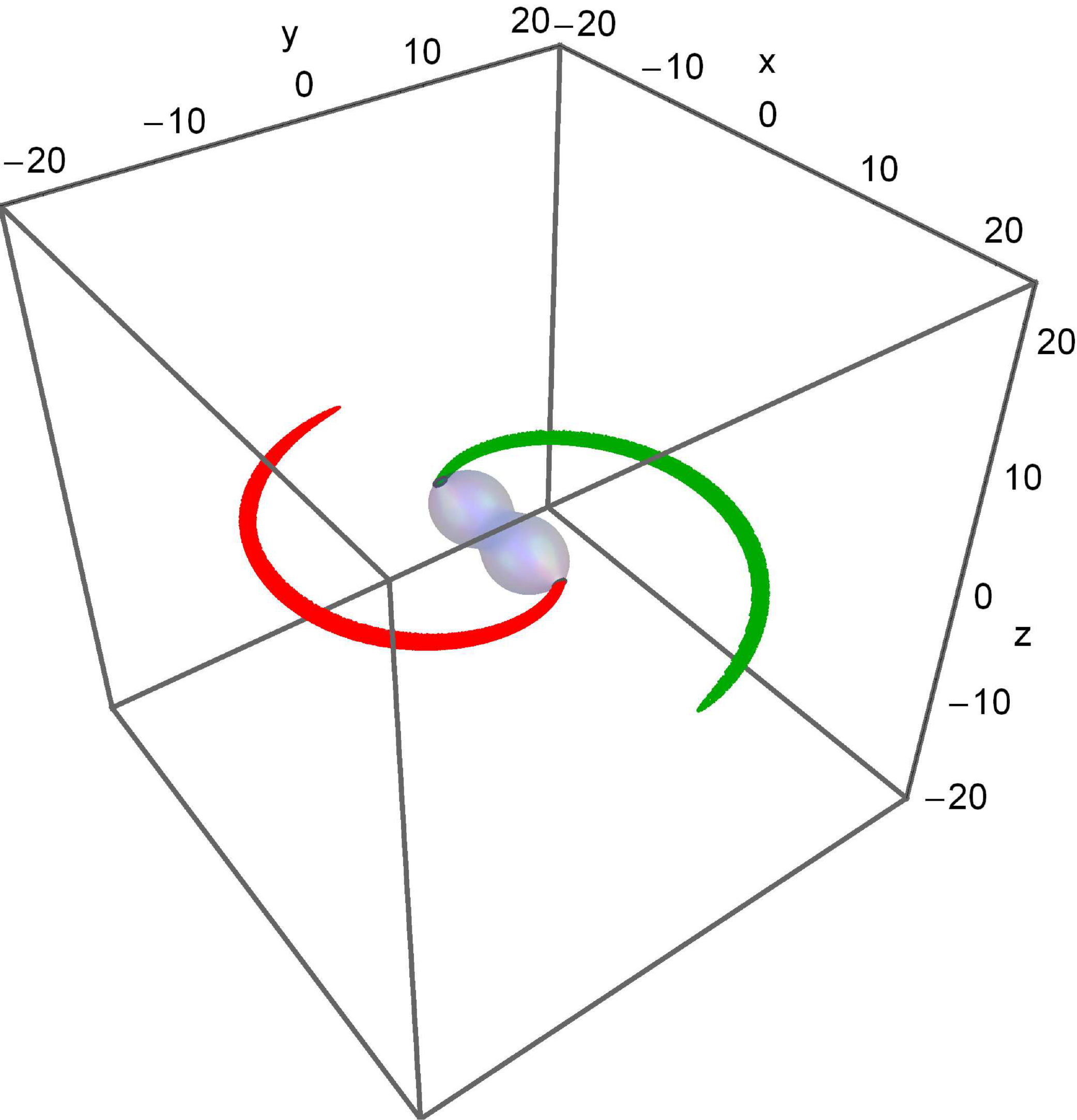}
\caption{The projection of the local segments of $W^u(NHIM)$ into the configuration $(x,y,z)$ space, when $E = -0.0575$. The manifolds from the NHIM over the saddle point $L_2$ are shown in red colour, while the ones from the symmetrically placed NHIM over $L_3$ are shown in green colour. The corresponding equipotential surface is shown in transparent gray colour. (For the interpretation of references to colour in this figure caption and the corresponding text, the reader is referred to the electronic version of the article.)}
\label{sprs}
\end{figure}

Stars which have an energy much higher than the saddle energy of the Lagrange points $L_2$ and $L_3$ will leave the system rather soon and for them the saddle dynamics described in this publication is not relevant. The situation is very different for stars with an energy just a little below this saddle energy. Occasionally, these stars suffer close encounters with other objects and thereby change their energy, they may gain a small amount of energy and afterwards have an energy just a little above the saddle energy. For exactly these stars the saddle dynamics is highly important.

After a while these stars approach one of the saddles $L_2$ or $L_3$ from the inside and along the local segment of the corresponding stable manifolds of the saddle NHIM. These stars remain for some time in the saddle region and then leave this region again along one of the branches of the unstable manifold of the saddle NHIM. Now it depends on the finest details of the initial conditions of the orbit whether the star leaves the saddle region to the inside of the system or to the outside. When it returns to the inside then later it will have another opportunity to come close to one of the saddles again. When it leaves to the outside then it is lost for the system and leaves to regions far away from the system and along the outer unstable manifolds of the saddle NHIM. Of course, such an escape is a rather rare event in total.

When an orbit starts in the vicinity of the saddle point $L_2$ then in forward direction (i.e. in the future) it converges automatically against the unstable manifold of the NHIMs ($W^u(NHIM)$). This observation provides the idea for a numerical construction of $W^u(NHIM)$. For this task we randomly select 5000 initial conditions close to $L_2$ all with the same energy $E = -0.0575$ and we let the orbits run for a finite time. Note that all these initial conditions start with $z_0 = 0$. In Fig. \ref{sprs} we plot the projection into the configuration $(x,y,z)$ space of the local segments of the unstable manifolds of the saddle NHIMs. The time interval used for the cut off is $t \in [0, 323]$\footnote{The final time of integration $t_{\rm fin} = 323$ time units was chosen so as the two spirals, emerging through the Lagrange points, to reach the opposite side, thus crating a half-circle path.}. In this figure, stars leaving the binary system over the saddle $L_2$ are plotted red, while stars leaving over the saddle $L_3$ are plotted green.

In the case of the binary system of the two dwarf galaxies the escaping stars run into empty space and therefore they do not trigger any spectacular observable effects. This is in big contrast to the barred galaxy, (e.g., \cite{JZ16}), where the few stars escaping over the saddles at the ends of the bar create visible stellar structures, such as spiral or ring patterns in the outer parts of the disk.

\section{Discussion}
\label{disc}

In this paper we numerically explored the escape dynamics of a 3-dof dynamical model describing a binary system of interacting dwarf spheroidal galaxies. Our study was focused on such energy levels which correspond to open equipotential surface with transport channels around the saddles $L_1$, $L_2$ and $L_3$. The test particles (stars) can travel either between the two primary bodies (dwarf galaxies), through $L_1$, or even enter the exterior region, through $L_2$ and $L_3$, and thus escape to infinity. We managed to classify initial conditions into bounded and escaping orbits, while the SALI method has been deployed for further distinguishing bounded motion into non-escaping regular as well as trapped sticky or chaotic motion. At the same time we located the various basins of escape, leading to different escape channels, while we also correlated them with the corresponding distribution of the escape time of the orbits. Our numerical analysis indicates that the escape mechanism of orbits in a binary galaxy system is a highly interesting, yet very complicated, procedure. The numerical values of the involved parameters (such as the masses, the distance between the two galaxies, etc) have been chosen having in mind the binary system of dwarf spheroidal galaxies NGC 147 and NGC 185.

A binary system of interacting galaxies is a very complex stellar entity and therefore we need to make some necessary assumptions and simplifications in order to be able to numerically study the orbital behaviour of the stars. In this vein, our model is rather simple and contrived to give us the ability to explore the different aspects of the dynamical model. Nevertheless, contrived models can always provide useful insight into more realistic stellar systems, which unfortunately are very difficult to be investigated, if we take into account all the astrophysical aspects. At this point, we must point out the main restrictions as well as the main limitations of our model: (i) The two dwarf galaxies are assumed to be coplanar, orbiting each other on circular orbits, (ii) our dynamical model deals only with the non-dissipative components of the galaxies, such as stars and/or possibly dark matter particles, (iii) the potentials we use are rigid and do not respond to the evolving density distribution. Therefore, the two galactic centers remain stationary in the rotating frame of reference. Moreover, since our galactic system is described by a conservative Hamiltonian (which implies that the total orbital energy of the stars does not change) it cannot model close encounters or collisions of stars.

Undoubtedly, the most important elements of the escape dynamics are the NHIMs, which control and direct the flow of particles over the saddle points of the effective potential. On this basis, the development scenario of the NHIMs, as a function of the total orbital energy, has been numerically investigated, along with the network of periodic orbits associated with the NHIMs. For studying the NHIMs we used the classical method of the restricted Poincar\'e map, by depicting the invariant surfaces in both two-dimensional and three-dimensional subspaces of the total six-dimensional phase space.

The NHIMs have their stable and unstable manifolds which play a crucial role on the escape dynamics. This is true because the manifolds are able to transport important dynamical structures, born in the vicinity of the NHIMs, far away from the central region of the potential. Consequently, the escaping orbits, which pass through the saddles, follow the unstable manifolds of the NHIMs, thus producing spiral arms projected into the configuration space. Our dynamical model predicts that spiral arms can be formed by escaping orbits through the saddles. However, around two dwarf galaxies there is no disk where the escaping orbits can trigger observable reactions (like the formation of new stars in barred galaxies when escaping stars run through the disk of gas and dust) and therefore there are no observations regarding spirals in binary systems. Nevertheless, at least theoretically the formation of spirals in binary systems is possible. The persistence of the NHIMs under perturbations (see e.g., \cite{F71} and  also chapter 3 in \cite{W94}) suggests that the scenario of spirals does not change qualitatively under small changes of the potential.

As far as we know this is the first detailed and systematic numerical exploration on the escape dynamics and the NHIMs of a 3-dof binary galaxy system and this is exactly the novelty and the contribution of our work. We hope that the present outcomes shed some light on the properties of the escape mechanism of stars and of course on the role of the NHIMs in binary galaxy systems. It is in our future plans to continue the numerical investigation of the system of two dwarf spheroidal galaxies in order to unveil the mysteries of the heteroclinic structures and the mutual influences of the NHIMs.

\section*{Acknowledgments}
One of the authors (CJ) thanks DGAPA for financial support under grant number IG-100616. We would like to express our warmest thanks to the two anonymous reviewers for the careful reading of the manuscript and for all the apt suggestions and comments which allowed us to improve both the quality and the clarity of our paper.

\begin{appendix}

\section{Computation of the initial value of the momenta along the $y$ axis}
\label{appex}

All orbits used for producing the colour-coded grids in Section \ref{esc} have pairs of initial conditions $(x_0,z_0)$, while $y_0 = p_{x0} = p_{z0} = 0$. Therefore, for a given energy level $E_0$, the Hamiltonian given in Eq. (\ref{ham}) reduces to
\begin{equation}
\frac{1}{2}p_{y0}^2 + \Phi_{\rm t}(x_0,z_0) - \Omega \ x_0 \ p_{y0} = E_0.
\label{ham2}
\end{equation}
The above equation can also be written in the form
\begin{equation}
\frac{1}{2}p_{y0}^2 - \Omega \ x_0 \ p_{y0} + \left(\Phi_{\rm t0} - E_0\right) = 0,
\label{quad}
\end{equation}
where $\Phi_{\rm t0} = \Phi_{\rm t}(x_0,z_0)$.

It is evident that Eq. (\ref{quad}) is in fact a quadratic equation, of the form $a p_{y0}^2 + b p_{y0} + c = 0$, with respect to $p_{y0}$, with
\begin{equation}
a = \frac{1}{2}, \ \
b = - \Omega \ x_0, \ \
c = \Phi_{\rm t0} - E_0.
\label{elm}
\end{equation}
The discriminant corresponding to Eq. (\ref{quad}) is
\begin{equation}
D = x_0^2 \Omega^2 - 2\left(\Phi_{\rm t0} - E_0\right).
\label{discr}
\end{equation}

Consequently the two roots are given by
\begin{equation}
p_{y0} = x_0 \ \Omega \pm \sqrt{x_0^2 \Omega^2 - 2\left(\Phi_{\rm t0} - E_0\right)}.
\label{rts}
\end{equation}
In our computations, regarding the colour-coded grids, we used the value of $p_{y0}$ with the plus (+) sign, in front of the square root.

\end{appendix}

\end{document}